\documentclass[graybox, envcountchap]{svmult}

\usepackage{mathptmx}        
\usepackage{amsmath}
\usepackage{amssymb}
\usepackage{color}
\usepackage{helvet}          
\usepackage{courier}         
\usepackage{dirtree}

\usepackage{makeidx}        
\usepackage{graphicx}        
\usepackage{subfig}

\usepackage{multicol}        
\usepackage[bottom]{footmisc}

\usepackage{hyperref}        
\hypersetup{colorlinks=true,urlcolor=blue}

\usepackage[misc]{ifsym}
\usepackage{esvect}

\makeindex             

\usepackage{xcolor}
\usepackage{ulem}

\begin{document}

\title{Tests of Classical Gravity with Radio Pulsars}
\author{Zexin Hu, Xueli Miao, Lijing Shao}
\authorrunning{Hu {\it et al.}}
\institute{Zexin Hu \at Department of Astronomy, School of Physics, 
Peking University, Beijing 100871, China \\
Kavli Institute for Astronomy and Astrophysics, 
Peking University, Beijing 100871, China \\ 
\email{huzexin@pku.edu.cn}
\and Xueli Miao \at National Astronomical Observatories, 
Chinese Academy of Sciences, Beijing 100012, China \\
\email{xlmiao@bao.ac.cn}
\and Lijing Shao (\Letter) \at Kavli Institute for Astronomy and Astrophysics, 
Peking University, Beijing 100871, China \\
National Astronomical Observatories, 
Chinese Academy of Sciences, Beijing 100012, China \\
\email{lshao@pku.edu.cn}}
%
%
\maketitle

\abstract{Tests of gravity are important to the development of our understanding
of gravitation and spacetime. Binary pulsars provide a superb playground for
testing gravity theories. In this chapter we pedagogically review the basics
behind pulsar observations and pulsar timing. We illustrate various recent
strong-field tests of the general relativity (GR) from the Hulse-Taylor pulsar
PSR~B1913+16, the double pulsar PSR~J0737$-$3039, and the triple pulsar
PSR~J0337+1715.  We also overview the inner structure of neutron stars (NSs)
that may influence some gravity tests, and have used the scalar-tensor gravity
and massive gravity theories as examples to demonstrate the usefulness of pulsar
timing in constraining specific modified gravity theories.  Outlooks to new
radio telescopes for pulsar timing and synergies with other strong-field gravity
tests are also presented.}


\section{Introduction}
\label{sec:intro}

As one of the four fundamental interactions, gravity is the most mysterious to
study.  From Newton's law of universal gravitation to Einstein's general
relativity (GR), in the development of gravity theory, experiments and
observations have been playing an important role in overturning or verifying
different gravity theories.  Currently, GR has passed all the tests with flying
colors~\cite{Will:2018bme}, and it seems that GR is the {\it tour de force}
among all gravity theories.  However, in GR, there is a fundamental difficulty
in quantization as well as in describing spacetime singularities, suggesting
that GR could not be a complete theory. It might be an effective field theory
below some energy scale~\cite{Weinberg:2021exr}.  So, there is strong motivation
to find deviations from GR predictions and to search for other possible
alternative gravity theories.  People have tried to construct quantum gravity
theories to solve the quantization problem in GR, such as the string
theory~\cite{Fradkin:1985ys}.  People have also used alternative gravity
theories to explain the missing mass in galaxies and the accelerating expansion
of our Universe without introducing the mysterious concepts of dark matter and
dark energy~\cite{Jain_2010, Clifton_2012}. To this end, we need to use all
kinds of observations to probe different aspects of gravity and use measurement
results to test gravity theories.

People usually classify gravity tests based on the studied systems in different
gravity regimes~\cite{Wex:2014nva, EventHorizonTelescope:2022xqj,
Kramer:2022gru}. In astrophysics, for systems of astronomical scales,
Wex~\cite{Wex:2014nva} introduced the following four gravity regimes: (i)
quasi-stationary weak-field regime, (ii) quasi-stationary strong-field regime,
(iii) highly-dynamical strong-field regime, and (iv) radiation regime.  In a
quasi-stationary weak-field regime, the velocity of masses is much less than the
speed of light $c$, and spacetime is close to a Minkowski spacetime. Our Solar
system belongs to this regime.  In a quasi-stationary strong-field regime, the
velocity of masses is also much less than $c$, but one or more masses  have
strong self-gravity, which causes the neighborhood spacetime to deviate
significantly from a Minkowski one.  A well-separated binary system containing
one or two compact bodies belongs to this regime.  In a highly-dynamical
strong-field regime, the velocity of masses is close to $c$ and spacetime is
strongly curved.  The merging of compact binaries reflects the gravity of such a
regime.  A radiation regime is the propagation regime of gravitational waves
(GWs).  Gravity tests from the Solar system have provided strict restrictions on
the quasi-stationary weak-field regime \cite{Will:2018bme}, while the GW
detection from LIGO/Virgo/KAGRA probes the radiation regime and provides
opportunities to analyze the highly-dynamical strong-field
regime~\cite{Sathyaprakash:2019yqt}.  To study the behavior of gravity in a
quasi-stationary strong-field regime, we need to have one or more compact bodies
in a well-separated system.  Binary pulsar  systems with good timing precision
provide precise measurements of gravity in such a regime.

In this chapter, we will provide a pedagogical introduction to pulsar timing, as
well as some aspects and new developments of using it in testing classic
(namely, not quantum) gravity theories.  In the next section, we overview the
basics of pulsar observations and the theoretical background for pulsar timing.
We use the famous Hulse-Taylor pulsar PSR~B1913+16, the double pulsar
PSR~J0737$-$3039, and the triple pulsar PSR~J0337+1715 as primary examples for
testing GR in Sec.~\ref{sec:ht:dp}. After that, we discuss the inner structure
of neutron stars (NSs) in Sec.~\ref{sec:ns} which in some theories will affect
the binary orbital dynamics. A couple of representative alternative gravity
theories are tested against pulsar timing data in Sec.~\ref{sec:timing}.
Finally, Sec.~\ref{sec:summary} summaries the chapter.  More reviews on the
topic of using radio pulsars for gravity tests can be found in
Refs.~\cite{Stairs:2003eg, Wex:2014nva, Manchester:2015mda, Shao:2016ezh,
Shao:2022izp, Kramer:2022gru}.


\section{Pulsar Observation and Pulsar Timing}
\label{sec:obsandtiming}

In 1967, the first pulsar PSR~B1919+21 was discovered~\cite{Hewish:1968Nature}.
The prevailing view of pulsars is that they are stably spinning NSs with a
high-intensity magnetic field~\cite{Gold:1968Natur}.  In a pulsar's open
magnetic field line region, the charged particles moving along magnetic field
lines can produce curvature radiation, creating a radiation cone centered on the
magnetic axis \cite{2004handbook}.  Generally, a pulsar's magnetic axis and spin
axis do not coincide, so if the radiation region of a NS sweeps past Earth's
radio telescopes when the NS rotates, one can detect a series of periodic pulse
signals on the Earth.  In this setting, a rotating NS seems like a lighthouse,
that is why we also call it the ``lighthouse model.'' Pulsars, as rapidly
rotating NSs, provide an ideal laboratory in astrophysics for fundamental
physics.  The masses of NSs are typically larger than that of our Sun, but their
radii are only about 10\,km. Therefore NSs represent an extremely dense stellar
environment.  Inside NSs there exists high matter density, which can reach and exceed 
nuclear density, $\sim 10^{15}\,{\rm g/cm^{3}}$, as well as high pressure, $\sim 10^{36}\,{\rm erg/cm^{3}}$. NSs also have high magnetic field
ranging from $10^{8}\,{\rm G}$ to $10^{15}\,{\rm G}$.

\begin{figure}[t]
	\centering
	\includegraphics[width=8.5cm]{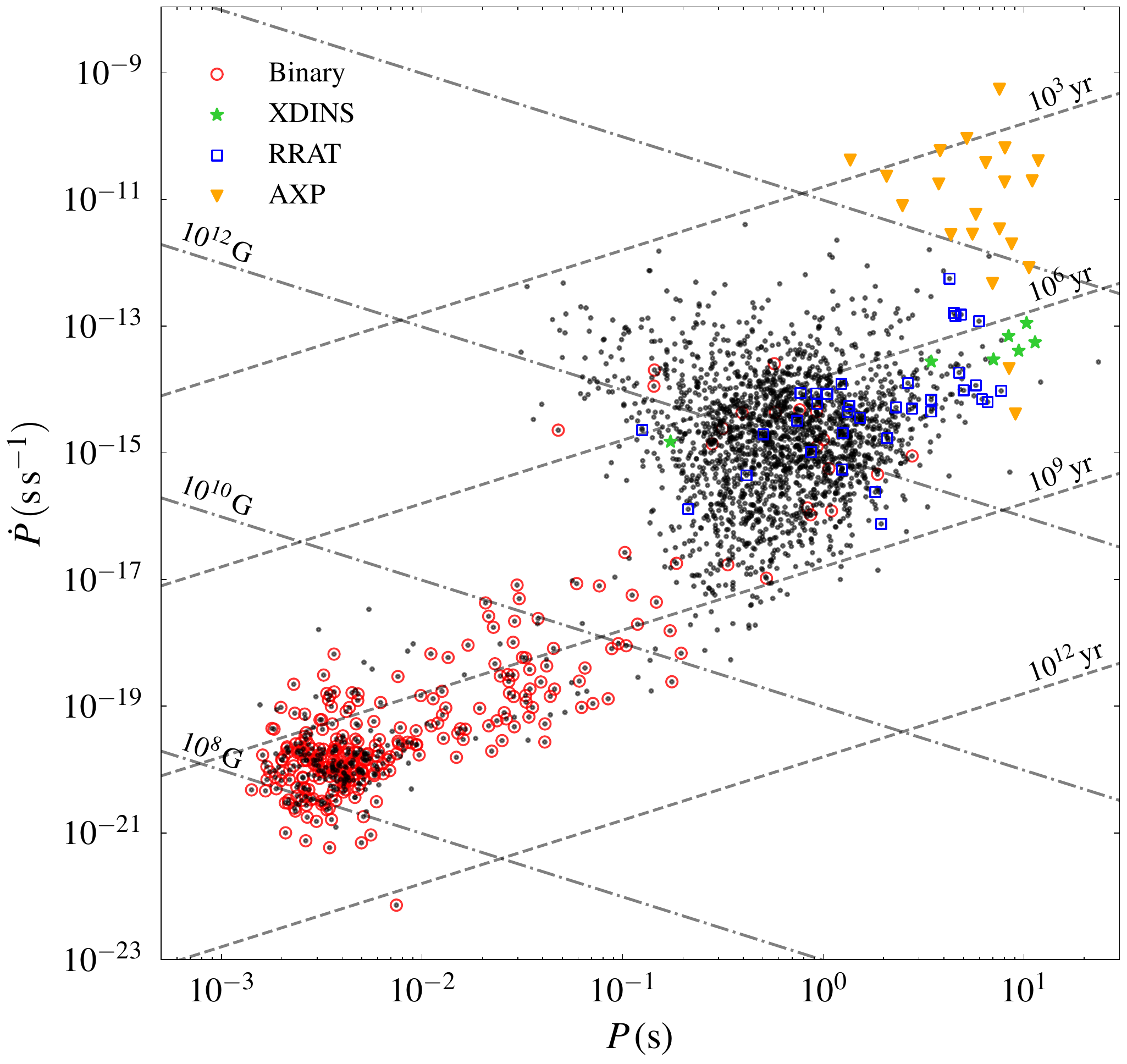}
	\caption{\label{fig:ppdot} The period--period derivative diagram for
	pulsars~\cite{Manchester:2005}. The black dots are pulsars with pulsed
	emission in the radio band and the red circles mark binary systems. We also
	denote ``AXP'' for systems that are anomalous X-ray pulsars or soft
	$\gamma$-ray repeaters, ``RRAT'' for systems that are pulsars with
	intermittently pulsed radio emission, and ``XDINS'' for systems that are
	isolated NSs with pulsed thermal X-ray emission but no detectable radio
	emission.}
\end{figure}

Benefiting from the development of telescopes, we have detected more than 3000
pulsars up to now.  Figure~\ref{fig:ppdot} shows the population of pulsars by
the distribution of spin period ($P$) and spin period derivative ($\dot{P}$) 
\cite{Manchester:2005}.  In the figure the red circle marks pulsars in binary
systems, which make up about $10\%$ of detected pulsars.  These pulsars' spin
periods range from $1.4\,{\rm ms}$ to $24\,{\rm s}$.\footnote{Recently, there are observations showing the existence of a NS with a spin period of 76\,{\rm s} \cite{Caleb:2022}.} According to the
distribution of $P$ and $\dot{P}$, one can classify pulsars into two groups: normal pulsars which correspond to the distribution in the upper right part in
Fig.~\ref{fig:ppdot}, and millisecond pulsars (MSPs) which correspond to the
distribution in the bottom left part.  For MSPs, as the name suggests, the spin
periods are mostly less than $10\,{\rm ms}$ and the spin period derivatives
range from $10^{-18}\, {\rm s\,s}^{-1}$ to $10^{-22}\, {\rm s\,s}^{-1}$.  The
small $\dot{P}$ of MSPs means that they have more stable spin periods than
normal pulsars on long timescales, and the stability of some MSPs can revive the
precision of atomic clocks in long timescales \cite{Lorimer:2005LRR,
Manchester:2005, Hobbs:2015}.  The fast rotation of MSPs is caused by the
``recycling process'' where the pulsars can spin up by a stable mass transfer
from the companion stars \cite{Tauris:2017ApJ}.  So, MSPs are mostly found in
binary systems, as clearly seen in Fig\,\ref{fig:ppdot}.  Fully recycled pulsars
would have millisecond spin periods and locate in very circular orbits. Some
pulsars could undergo mild recycling processes and only spin up to $\gtrsim
10$\,ms, and their orbits would be eccentric. The stability of spin periods of
MSPs makes the measurements of them highly precise. Detecting a stably 
rotating MSP in a binary system allows us to test gravity in a quasi-stationary 
strong-field regime.

The radiation of a pulsar can be detected in multiple frequency bands.  For
example, {the pulses from} the Crab pulsar, a young pulsar discovered 
in the Crab nebula, { have been detected} from the radio 
bands to $\gamma$-ray bands \cite{1968Sci_stealin,
Lundgren:1995ApJ}.  The discovery of the Crab pulsar in the Crab nebula also
demonstrates that a rotating NS  is a product of a supernova.  While other bands
also provide important information, study of pulsars is mainly benefited from
the radio band. Pulsars usually have a power-law radio spectrum and the spectrum
index is normally from $-2$ to $-1.5$ \cite{2004handbook}.

The pulse radiation passes through the ionized interstellar medium (ISM) before
being detected by radio telescopes. The ISM causes radio signals to disperse in
the propagation, so the timing of the signal is frequency dependent. For signal
of frequency $f$, the time delay $t$ caused by ISM follows $t = \textit{D}\times
{\rm DM}/f^{2}$, namely that the signals of higher frequency arrive earlier than
those of lower frequency.  The parameter $\textit{D}\equiv{e^{2}}/{2\pi m_{e}c} =
(4148.808\pm0.003) \,{\rm MHz^{2} \, pc^{-1} \, cm^{3} \, s}$ is the dispersion
constant, where $e$ and $m_{e}$ are the charge and mass of an electron
respectively.  ``DM'' is the dispersion measure, which controls the magnitude of
the delay, defined via ${\rm DM} \equiv \int^{d}_{0}n_{e}{\rm d}l$, where
$d$ is the distance of the pulsar and $n_{e}$ is the electron number density.  Using
the time delay between two different frequencies ($f_1$ and $f_2$), 
\begin{equation}
    {\rm \Delta} t = \textit{D}\times \big(f_{1}^{-2}-f_{2}^{-2} \big)\times{\rm DM}\,,
\end{equation}
we can calculate the value of DM.  With a Galactic electron density distribution
model, we can calculate the distance of the pulsar.  There are two mainstream
models, the NE2001 model \cite{Cordes:2002astro} and the YMW model
\cite{Yao:2017ApJ}. As post processing for the observation, to rectify the dispersion effect during
propagation, one can adopt ``incoherent dedispersion'' or ``coherent
dedispersion'' techniques.

In reality, the ISM is highly turbulent and inhomogeneous, and the irregularity
induces extra effects on the propagating signals. It causes multi-path
scattering of the radio signals, leading pulse profiles of pulsars to have
``exponential tails'' which blur the profile edges and worsen the timing
precision. The scattering effect decreases with frequency  roughly as $f^{-4}$,
so using a higher frequency band to observe reduces the scattering effect
\cite{2004handbook}.  In addition, the relative motion between the pulsar, the
scattering screen, and the radio telescope can cause interstellar scintillation.
It leads to intensity variations of the pulsar's signals on various timescales
and frequencies \cite{2004handbook}.

\begin{figure}[t]
	\centering
	\includegraphics[width=6.6cm]{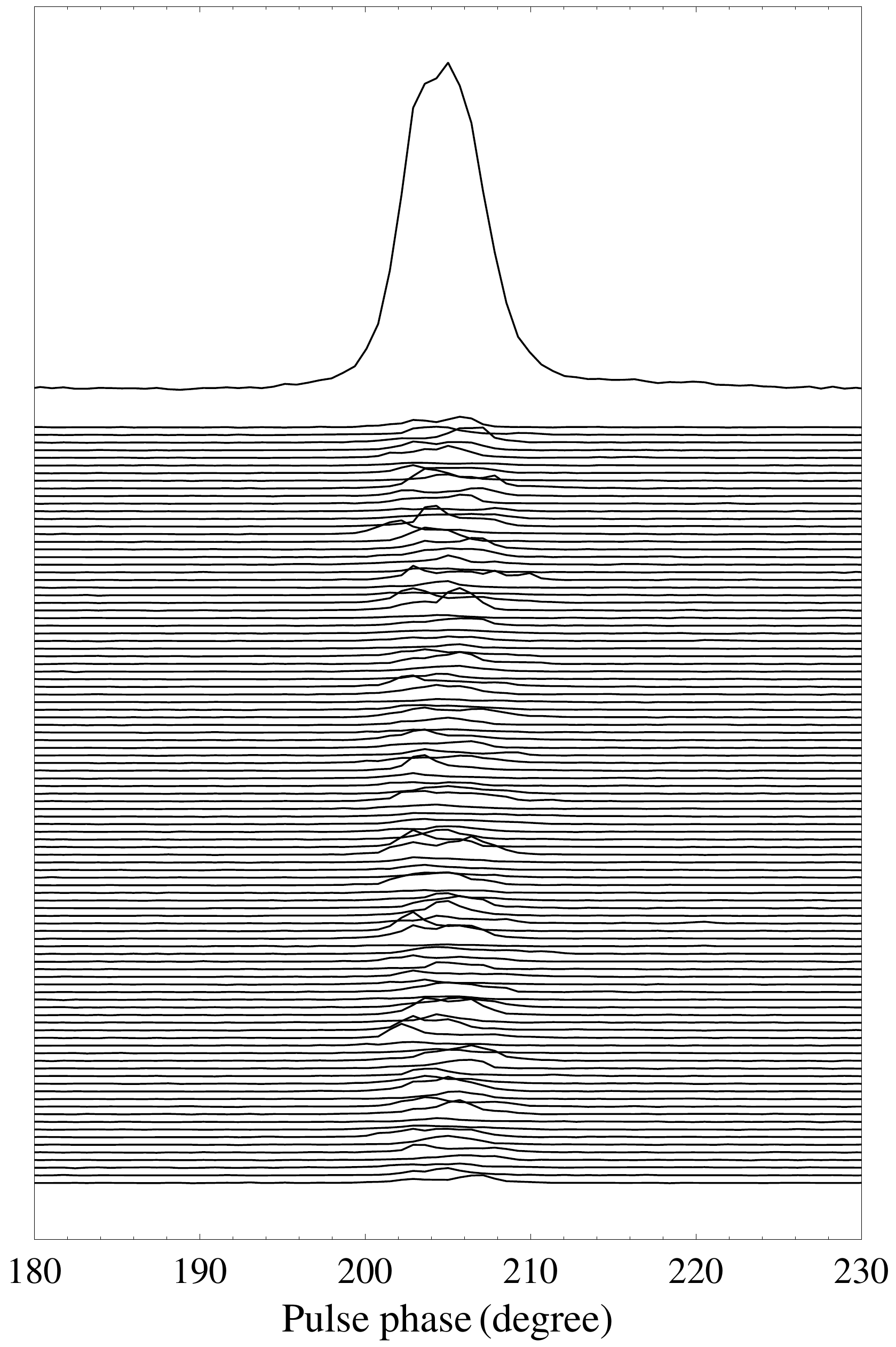}
	\caption{\label{fig:singlepulse} Stack of 100 single pulses from PSR
	J2222$-$0137, observed with the FAST telescope \cite{Miao:2023gkr}. The  line in the top is the
	integrated profile of these 100 pulses.}
\end{figure}

After accounting for the dispersion effects of pulse signals, we get a series of
single pulses.  Figure~\ref{fig:singlepulse} shows 100 continuous single pulses
versus spin phase from PSR~J2222$-$0137, which was observed by the
Five-hundred-meter Aperture Spherical radio Telescope (FAST).  In the figure,
the pulse-to-pulse variation of profiles is not stable.  Generally, the profiles
of single pulses of a pulsar are stochastic.  But some pulsars exhibit unique
temporal characteristics, such as sub-pulse drifting \cite{1968Natur_Drake,
2006A_A_Weltevrede}, mode changing \cite{1982ApJ_Bartel}, nulling
\cite{1970Natur_backer, 2007MNRAS_Wang}, giant pulse \cite{1968Sci_stealin,
2003Nature_hankins}, and microstructures \cite{1990AJ_Cordes}.  The
investigation of pulsars' single pulses helps one reveal the radiation mechanism
of pulsars and the structure of the emission region. Here we do not provide a
detailed introduction on the single-pulse study, more relevant studies on this topic can be found in Refs.~\cite{Shannon:2012tr,Liu:2016bae}.  We will concentrate on
introducing the pulsar timing technique that helps us  measure pulsars'
physical properties.

For pulsar timing, we measure the times of arrival (TOAs) of a pulsar's radio
signals at radio telescopes on the Earth and use a timing model to
fit the TOAs to get a phase-connection solution \cite{Damour:1991rd,
Wex:2014nva}.  To generate TOAs, we should use pulse profiles to cross correlate
with the ``standard profile'' of the pulsar at that observing frequency.  To
achieve high-precision TOAs, we need to use stable profiles of pulses.  The
shape of  profiles is highly variable from pulse to pulse. The integration of
hundreds or thousands of pulse periods leads to stable profiles.  In addition,
except for bright pulsars, single pulses of a pulsar are faint, and it is
difficult to detect single pulses. So, to get a stable profile with a high
signal-to-noise ratio (SNR), we need to fold hundreds or thousands of pulse
signals according to a pulsar's spin period. In Fig.~\ref{fig:singlepulse}, the
profile in the top is the integrated profile of PSR~J2222$-$0137, and it is
folded from the 100 single pulses \cite{Miao:2023gkr}.  Generally, we cannot simply get the spin
period directly from the pulse TOA intervals. The propagation effects on signals
and the motion of pulsars cause various time delays to TOAs. We need to adopt a
specific timing model to predict these time-delay effects and get the intrinsic
rotation periods.

The intrinsic rotation frequency of a pulsar can be Taylor-expanded by
\cite{Damour:1991rd, 2004handbook},
\begin{equation}\label{eq:spinfrequency}
    \nu(T)=\nu_{0}+\dot{\nu}_{0}\left(T-t_{0}\right)+\frac{1}{2} \ddot{\nu}_{0}\left(T-t_{0}\right)^{2}+\cdots\,,
\end{equation}
where $T$ is pulsar proper time, $\nu_{0} \equiv \nu(t_{0})$ is the spin
frequency at reference epoch $t_{0}$, $\dot{\nu}_{0}$ and $\ddot{\nu}_{0}$ are
the first and second time derivatives of spin frequency at $t_0$.  In the
magnetic dipole model, the radiation of a pulsar takes away the rotation energy
and makes the pulsar spin down, which contributes to $\dot{\nu}_{0}$ and
$\ddot{\nu}_{0}$.  We can also rewrite Eq.~(\ref{eq:spinfrequency})  as the
relation between the pulsar's proper time and pulse number
$N$~\cite{Damour:1991rd},
\begin{equation}\label{eq:pulsenumber}
  N=N_{0}+\nu_{0}\left(T-t_{0}\right)+\frac{1}{2} \dot{\nu}_0 \left(T-t_{0}\right)^{2}+\frac{1}{6} \ddot{\nu}_0\left(T-t_{0}\right)^{3}+\cdots\,.
\end{equation}

Equations~(\ref{eq:spinfrequency}) and (\ref{eq:pulsenumber}) describe the
pulsar rotation in a reference frame co-moving with the pulsar.  For us, what we
measure is TOAs of pulses recorded by an atomic clock on the Earth.  The
recorded TOAs are in the non-inertial frame of reference where the radio
telescope is located.  So we need to translate the TOAs to an inertial frame,
and the Solar System Barycentre (SSB) is an excellent approximate inertial
reference frame.  For a solitary pulsar, its proper time $T$ equals to its time
at SSB of infinite frequency up to a constant. It can be expressed by
\cite{Damour:1991rd},
\begin{equation}\label{eq:solarsystemdelay}
  T = t_{\rm tele}+t_{c}- \frac{D \times {\rm DM}}{f^{2}}+{\rm \Delta}_{\rm R_{\odot}}+{\rm \Delta}_{\rm S_{\odot}}+{\rm \Delta}_{\rm E_{\odot}}\,,
\end{equation}
where $t_{\rm tele}$ is the time that a pulse arrivals at a telescope, $t_{c}$
is the relative correction between an atomic clock of the telescope to an
average reference clock, $D \times {\rm DM} /f^{2}$ is the correction of
dispersion effect, $f$ is the observing frequency, ${\rm \Delta}_{\rm
R_{\odot}}$ is the R$\ddot{\rm o}$mer delay caused by the motion of the Earth
around the Sun, ${\rm \Delta}_{\rm S_{\odot}}$ is the Shapiro delay which is a
relativistic delay from the curvature of spacetime caused by the presence of
masses in the Solar system (notably the Sun and the Jupiter), ${\rm \Delta}_{\rm
E_{\odot}}$ is the Einstein delay which is a combined effect of gravitational
redshift caused by masses in the Solar system and time dilation caused by the
motion of the Earth.

In gravity tests with radio pulsars, binary pulsar systems are the main objects
of investigation.  For a pulsar in a binary system, the orbital motion of the
pulsar and the gravity field of the companion lead to extra time delays.
For binary pulsar systems, Eq.~(\ref{eq:solarsystemdelay}) is extended to
\cite{Damour:1991rd},
\begin{align}\label{eq:binarymodel}
     T=t_{\rm tele}+t_{c}- \frac{D \times {\rm DM}}{f^{2}}+{\rm \Delta}_{\rm R_{\odot}}+{\rm \Delta}_{\rm S_{\odot}}+{\rm \Delta}_{\rm E_{\odot}}+{\rm \Delta}_{\mathrm{RB}}+{\rm \Delta}_{\mathrm{SB}}+{\rm \Delta}_{\mathrm{EB}}+{\rm \Delta}_{\mathrm{AB}}\,.
\end{align}
Compared with Eq.~(\ref{eq:solarsystemdelay}), it has four additional terms:
${\rm \Delta}_{\mathrm{RB}}$ is the R$\ddot{\rm o}$mer delay caused by the
orbital motion of the pulsar; ${\rm \Delta}_{\mathrm{SB}}$ is the Shapiro delay
caused by the gravity field of the companion star in the binary; ${\rm
\Delta}_{\mathrm{EB}}$ is the Einstein delay which is caused by the time
dilation and gravitational redshift, and they are due to the orbital motion of
the pulsar; ${\rm \Delta}_{\mathrm{AB}}$ is the aberration delay that results
from aberration of the rotating pulsar beam, and this effect depends on the
time-varying transverse component of the orbital velocity of the pulsar
\cite{2004handbook}.

The motion of a pulsar in a binary orbit can be described by six Keplerian
parameters, and they are the orbital period of the pulsar $P_{b}$, the orbital
eccentricity $e$, the projected semi-major axis $x$, the longitude of periastron
relative to the ascending node $\omega$, the time of periastron passage $T_{0}$,
and the position angle of the ascending node $\Omega_{\rm asc}$ which we
generally cannot measure solely from pulsar timing unless the binary pulsar is
very nearby \cite{2004handbook}.  When the six Keplerian parameters are
determined, we can describe the motion of a binary system well.  However, these
parameters change with time because of various reasons. For example, for a
binary pulsar system, the pulsar and the companion are slowly inspiraling and
the system radiates GWs outward. So we need to consider orbital evolution over
time.

After the first binary pulsar system PSR~B1913+16 was detected
\cite{Hulse:1975ApJ}, how to provide a simple timing model with correct binary
motion became a hot topic that was constantly discussed at that time.
Considering the first-order post-Newtonian (PN) approximation and combining with
the pulsar observational data, Damour and Deruelle \cite{Damour1985,Damour1986}
proposed a phenomenological parametrization---the so-called parametrized
post-Keplerian (PPK) formalism---and later Damour and Taylor
\cite{Damour:1991rd} extended it.  The PPK parameters reflect effects
beyond the Keplerian orbit that can be extracted from pulsar timing and do not
depend on specific boost-invariant gravity theories at the 1\,PN order.  The
Damour-Deruelle (DD) timing model gets the orbital motion information by fitting
a series of Keplerian and PPK parameters.  The commonly used observable PPK
parameters are
\cite{Damour1986,Damour:1991rd},
\begin{equation}\label{eq:pkparameters}
    \{p^{\rm PPK}\}=\{\dot{\omega},\,\gamma,\,\dot{P}_{b},\,r,\,s,\,\delta_{\theta}\}\,,
\end{equation}
where $\dot{\omega}$ is the advance rate of periastron, $\dot{P}_{b}$ is the
decay rate of orbital period, $\gamma$ is the amplitude of ${\rm
\Delta}_{\mathrm{EB}}$, $r$ and $s$ are the PPK parameters describing ${\rm
\Delta}_{\mathrm{SB}}$, $\delta_{\theta}$ is a PPK parameter in ${\rm
\Delta}_{\mathrm{RB}}$. 

Considering GW radiation makes orbital parameters change with time.  Based on
the Kepler's equation, Damour and Deruelle \cite{Damour1985, Damour1986}
provided a Kepler-like equation,
\begin{equation}\label{eq:qusikepler}
    u-e \sin u= \frac{2\pi}{P_{b}}\left[\left(T-T_{0}\right)-\frac{\dot{P}_{b}}{2} \frac{\left(T-T_{0}\right)^{2}}{P_{b}}\right]\,,
\end{equation}
where $u$ is the eccentric anomaly, and its relation to the true anomaly
$A_{e}(u)$ is,
\begin{equation}
    A_{e}(u)=2\arctan{\left[\sqrt{\frac{1+e}{1-e}} \tan \frac{1}{2} u\right]}\,.
\end{equation}
The longitude of periastron $\omega$ can be modified as, $\omega=\omega_{0}+
{\dot{\omega}}A_{e}(u) / {n_{b}}$, where $n_{b}=2\pi/P_{b}$ is the average
angular velocity of the orbit.

Damour and Deruelle  \cite{Damour1986} provided the time delay terms for binary
systems. The R$\ddot{\rm o}$mer delay is,
\begin{align}\label{eq:romer}
{\rm \Delta}_{\rm RB}= x \sin \omega\left[\cos u-e\left(1+\delta_{r}\right)\right] 
+x\left[1-e^{2}\left(1+\delta_{\theta}\right)^{2}\right]^{1 / 2} \cos \omega \sin u\,,
\end{align}
where $\delta_{r}$ is also a PPK parameter, but we cannot get it by directly
fitting it in the timing model, because it can be absorbed by redefining other
parameters.  $\delta_{\theta}$ and $\delta_{r}$ describe the relativistic
deformations of the orbit. Damour and Deruelle used three eccentricities, and
the two new eccentricities are defined via $e_{\mathrm{r}} \equiv
e\left(1+\delta_{\mathrm{r}}\right)$ and $e_{\theta} \equiv
e\left(1+\delta_{\theta}\right)$.

The Einstein delay of a binary is,
\begin{equation}\label{eq:einstein}
    {\rm \Delta}_{\rm EB}=\gamma\sin{u}\,,
\end{equation}
where $\gamma$ represents the amplitude combining  time dilation and
gravitational redshift.  The Einstein delay is degenerate with the R$\ddot{\rm
o}$mer delay, and the degree of degeneracy depends on the change of $\omega$
\cite{Damour:1991rd}, namely, if $\omega$ changes significantly, the degeneracy
can be broken.

The Shapiro delay of a binary is~\cite{Blandford:1976},
\begin{equation}\label{eq:Shapiro}
{\rm \Delta}_{\rm SB}=-2 r \ln \left\{1-e \cos u-s\left[ \sin \omega(\cos u-e)+\left(1-e^{2}\right)^{1 / 2} \cos \omega \sin u\right]\right\}\,,
\end{equation}
where $r$ and $s$ are the above-mentioned range and shape parameters
respectively.  The Shapiro delay effect will be noticeable when the orbital
inclination $i$, which is the inclination of the orbit with respect to the line
of sight, is close to $90^\circ$, that is, the binary system is close to
edge-on.  Otherwise, this effect can be partially absorbed in other timing
parameters.  A strong Shapiro delay effect provides a direct measurement of
the mass of the companion and the inclination angle $i$.  For low-eccentricity
binary systems, the PPK parameters $r$ and $s$ have a high correlation,
especially when the inclination angle $i$ is not sufficiently  close to
$90^\circ$.  In this situation, we cannot get $r$ and $s$ with high precision. 
Freire and Wex \cite{Freire:2010mnras} provided two new PPK parameters, $h_{3}$
and $h_{4}$, to alleviate this problem.  The PPK parameters $h_{3}$ and $h_{4}$
are the amplitudes of the third and fourth harmonics from the Shapiro delay's
Fourier expansion.  Compared with $r$ and $s$, $h_{3}$ and $h_{4}$ are less
correlated with each other.  For edge-on systems,  parameters $h_{3}$ and
$\zeta$ are introduced to replace $h_{3}$ and $h_{4}$, where $\zeta\equiv
h_{3}/h_{4}$ is the ratio of amplitudes of successive harmonics.  The parameters
$h_{3}$ and $\zeta$ provide a superior description of the constraints on the
orbital inclination $i$ and the masses of the binary than $r$ and $s$.  In the
new parametrization, the Shapiro delay is rewritten as \cite{Freire:2010mnras},
\begin{align}
{\rm \Delta}_{\mathrm{SB}}=-\frac{2 h_3}{\zeta^3} \ln \left(1+\zeta^2-2 \zeta \sin \Phi\right) \,,
\end{align}
where $\Phi$ is the longitude relative to the ascending node.  The relations
between $(h_{3},\,\zeta)$ and $(s,\,r)$ are $s = {2\zeta}/({\zeta^{2} + 1})$ and
$r = {h_{3}}/{\zeta^{3}}$ \cite{Freire:2010mnras}.

Finally, the aberration delay is \cite{2004handbook},
\begin{equation}\label{eq:aberration}
{\rm \Delta}_{\rm{AB}}=A\left\{\sin \left[\omega+A_{e}(u)\right]+e \sin \omega\right\}+B\left\{\cos \left[\omega+A_{e}(u)\right]+e \cos \omega\right\}\,,
\end{equation}
where $A$ and $B$ are both PPK parameters, but they are not separately
measurable by pulsar timing.

After considering these time delay terms contributed to the timing model, we can
precisely measure orbital parameters by pulsar timing.  For boost-invariant
gravity theories, in the DD timing model, the PPK parameters are  functions of
Keplerian parameters, masses of the binary and parameters of gravity theories.
In GR, PPK parameters are
functions of Keplerian parameters and masses of the binary only. In GR, PPK
parameters in Eq.~(\ref{eq:pkparameters}) are expressed as
\cite{2004handbook,Wex:2014nva},
\begin{align}
  \dot{\omega} &= \frac{3 n_b}{1-e^2} \frac{V_b^2}{c^2}\,, \label{eq:omegadot}\\
   \gamma & =\frac{e}{n_b}\left(1+\frac{m_c}{m}\right) \frac{m_c}{m} \frac{V_b^2}{c^2}\,, \label{eq:gamma}\\
   r & =\frac{G m_c}{c^3}\,, \label{eq:r}\\
   s & =\sin{i}=x n_b \frac{m}{m_c} \frac{c}{V_b} \,, \label{eq:s}\\
  \delta_{\theta}&=\frac{({7}/{2}) m_{p}^{2}+6 m_{p} m_{c}+2 m_{c}^{2}}{m^{2}}\frac{V_{b}^{2}}{c^{2}} \,,\label{eq:deltatheta}\\
    \dot{P}_b&=-\frac{192 \pi}{5} \frac{m_p m_c}{m^2} f(e) \frac{V_b^5}{c^5}\,, \label{eq:pbdot}
\end{align}
where $V_{b} \equiv \left(Gmn_{b} \right)^{1/3}$ and
$f(e)={\big[1+(73/24)e^{2}+(37/96)e^{4}\big]}{(1-e^{2})^{-7/2}}$.  In the above
equations, $m_{p}$ is the mass of the pulsar, $m_{c}$ is the mass of the companion, $m$
is the total mass of the binary and $x \equiv a_{p}\sin{i}/c$ is the projected
semi-major axis of pulsar's orbit in the unit of seconds.

In relativistic gravity theories, the spins of the bodies that orbit in a binary
system can affect their orbital and spin dynamics.  Three contributions are
induced by a rotating body moving in a binary pulsar system at the leading
orders.  They are the spin-orbit interaction between the spin of pulsar
$\textbf{S}_{p}$ and the orbital angular momentum $\textbf{L}$, the spin-orbit
interaction between the spin of companion $\textbf{S}_{c}$ and $\textbf{L}$, and
the spin-spin interaction between $\textbf{S}_{p}$ and $\textbf{S}_{c}$
\cite{Wex:2014nva}.  The spin-spin interaction is generally many orders of
magnitude smaller than the 2\,PN terms and the spin-orbit effects, so that one
can ignore it. We only introduce the spin-orbit interaction here. In a binary
pulsar system, the spin-orbit interaction leads the orbit and the two spins to
precess \cite{Wex:2014nva,Bagchi_2018}. The main effect of the precession of the
spin is a secular change in the orientation of the spin, which is caused by the
effect of spacetime curvature and the precessing rate is independent of the
spin.  This effect is the so-called geodetic precession, and it has been
detected in some binary pulsar systems, such as PSR~J0737$-$3039
\cite{Breton2008} and PSR~J1906+0746 \cite{Desvignes:2019uxs}. 

For a pulsar in a binary system, geodetic precession causes the pulsar's spin to
precess around the total angular momentum. As the orbital angular momentum is
normally much larger than the pulsar's spin, we can regard that the pulsar's
spin precesses around the orbital angular momentum.  In GR, the average rate of
geodetic precession of the pulsar's spin at the leading order is
\cite{2004handbook}, 
\begin{equation}
    \Omega_{p}=\frac{n_b}{1-e^2}\left(2+\frac{3 m_c}{2 m_p}\right) \frac{m_p m_c}{m^2} \frac{V_b^2}{c^2}\,.
\end{equation}
The secular change of the spin orientation causes changes in the observed emission
properties of the pulsar because the line-of-sight will cut through different
regions of the magnetosphere.  One can detect a changing integrated profile and polarization angle (PA) in data. In turn,
one can use the change of profile or PA to calculate ${\rm \Omega}_{p}$, and
compare the results with GR's predictions~\cite{Desvignes:2019uxs}.

 The orbital precession 
is mainly contributed by two effects.  The first
effect is  the mass-mass interaction, which belongs to the PN terms of two
point masses.  The second effect is  the Lense-Thirring effect related to
the spins of two stars \cite{Bagchi_2018}.  Damour and
Schaefer~\cite{Damour:1988} decomposed the precession of the orbit in terms of
observable parameters of pulsar timing. One of the parameters is
$\dot{\omega}_{p}$, which can be decomposed into,
\begin{align}\label{eq:omegalense}
    \dot{\omega}_{p}=\dot{\omega}_{\rm PN}+\dot{\omega}_{{\rm LT}_{p}}+\dot{\omega}_{{\rm LT}_{c}}\,,
\end{align}
where $\dot{\omega}_{\rm PN}$ is the PN term and the 1\,PN order is given in
Eq.~(\ref{eq:omegadot}); $\dot{\omega}_{{\rm LT}_{p}}$ and $\dot{\omega}_{{\rm
LT}_{c}}$ are due to the Lense-Thirring effect of the pulsar and the companion
respectively.  For a binary NS system, the spin of the companion is generally slow,
so the $\dot{\omega}_{{\rm LT}_c}$ that comes from the contribution of the
companion's spin can be ignored.  If $\textbf{S}_{p}$ is parallel to the orbital
angular momentum $\textbf{L}$, the $\dot{\omega}_{{\rm LT}_{p}}$ can be
simplified to \cite{Bagchi_2018, Hu:2020ubl},
\begin{align}
    &\dot{\omega}_{{\rm LT}_{p}} = -\frac{3n_{b} }{1-e^{2}} \beta^{\rm s}_{p} \,
    g_{p\,\parallel}^{\rm s} \, \frac{V_{b}^{3}}{c^{3}} \,, \label{eq:omdotLT}
\end{align}
where $X_{p} \equiv m_{p}/m$, $\beta^{\rm s}_{p} \equiv {cI_{p}}n_{b}/
Gm^{2}_{p}$, $g_{p\,\parallel}^{\rm s} \equiv {\left(\frac{1}{3} X^{2}_{p}+
X_{p} \right)}{ \left( 1-e^{2} \right)^{-1/2}}$, and $I_{p}$ is the moment of
inertia of the pulsar which depends on the equation of state (EOS) of NSs. As we
can see, the value of $\dot{\omega}_{{\rm LT}_{p}} $ depends on $I_{p}$. So if
we can measure $\dot{\omega}_{{\rm LT}_{p}} $, we can get a value of $I_{p}$
which can help us determine the EOS of NSs.  However, using the best-measured
double pulsar as an example~\cite{Kramer:2009zza, Hu:2020ubl, Kramer:2021jcw},
the value of $\dot{\omega}_{{\rm LT}_{p}}$ is at the same order as the value
of $\dot{\omega}_{\rm 2\,PN}$, so only highly relativistic binary pulsar systems
with high-precision timing results might have the ability to limit the moment of
inertia of NSs with this method.

\section{Hulse-Taylor Pulsar, Double Pulsar, and Triple Pulsar}
\label{sec:ht:dp}

With the theoretical background in the last section, we now introduce three of
the most famous radio pulsars in the field, the Hulse-Taylor pulsar PSR~B1913+16
\cite{Hulse:1975ApJ, 1979Natur_Taylor, 1913_2016ApJ}, the double pulsar
PSR~J0737$-$3039A/B \cite{Burgay:2003Natur, Lyne:2004Sci, Kramer:2006Sci,
Breton2008, Kramer:2021jcw}, and the triple pulsar
PSR~J0337+1715~\cite{Ransom:2014, Archibald:2018oxs, Voisin:2020lqi}.

\subsection{The first binary pulsar: PSR~B1913+16}

PSR~B1913+16, a recycled pulsar in a highly eccentric 7.75-hr orbit, was the
first pulsar detected in a binary system. The binary is a double NS system, and
it was discovered using the Arecibo telescope in 1974 \cite{Hulse:1975ApJ}.  We
also call this pulsar the Hulse-Taylor pulsar in honor of its discoverers.  The
 observation of this system indirectly proved the existence of GW radiation for the first time
\cite{1979Natur_Taylor}.  The latest timing results of this pulsar are published
in Ref.~\cite{1913_2016ApJ}, where 9257 TOAs were acquired over 35 years, and
the analysis provided  stringent tests of gravity theories. The orbital
parameters in DD timing model are listed in Table~\ref{tab:PSR1913}. Three PPK
parameters---the advance rate of periastron $\dot{\omega}$, the amplitude of
Einstein delay $\gamma$, and the decay rate of orbital period
$\dot{P}_{b}$---were well determined~\cite{1913_2016ApJ}. In addition, two PPK
parameters, $r$ and $s$, of the Shapiro delay effect from this system were
measured. All these measurements are consistent with GR predictions.  The
relativistic shape correction to the elliptical orbit, $\delta_{\theta}^{\rm
obs}$, is also detected for the first time.  

\begingroup
\setlength{\tabcolsep}{10pt}
\renewcommand{\arraystretch}{1.2}
\begin{table}[t]
\begin{center}
\caption{Orbital parameters of PSR B1913+16 \cite{1913_2016ApJ}.}
\label{tab:PSR1913}       
\begin{tabular}{p{4cm}p{3cm}}
\noalign{\smallskip}\svhline\noalign{\smallskip}
Parameter & Value  \\
\hline\noalign{\smallskip}
$x=a_{p}\sin{i}/c ~ ({\rm s})$ & 2.341776(2)\\
$e$ & 0.6171340(4)\\
$P_{b} ~ ({\rm d})$  & 0.322997448918(3)\\
$\omega_0 ~ ({\rm deg})$  & 292.54450(8)\\
$\dot{\omega} ~ ({\rm deg\,yr^{-1}})$ & 4.226585(4)\\
$\gamma ~ ({\rm ms})$  & 0.004307(4)\\
$\dot{P}_{b}^{\rm obs}$ & $ -2.423(1) \times 10^{-12}$ \\
$\delta_{\theta}^{\rm obs}$ & $ 4.0(25) \times 10^{-6}$ \\
$s$ & $0.68^{+0.10}_{-0.06}$ \\
$r$ $(\mu{\rm s})$ & $9.6^{+2.7}_{-3.5}$ \\
\noalign{\smallskip}\hline\noalign{\smallskip}
$\zeta$ & $0.38(4)$ \\
$h_{3}$ $({\rm s})$ & $0.6(1)\times10^{-6}$ \\
\noalign{\smallskip}\hline\noalign{\smallskip}
\end{tabular}
\end{center}
\end{table}
\endgroup

Assuming GR is the true theory of gravity,
Eqs.~(\ref{eq:omegadot}--\ref{eq:gamma}) were used to get the masses of the
binary, $m_{p}=1.438\pm0.001\,M_{\odot}$ and
$m_{c}=1.390\pm0.001\,M_{\odot}$~\cite{1913_2016ApJ}. Inserting the measured
Keplerian parameters and the derived masses into Eq.~(\ref{eq:pbdot}), Weisberg
and Huang~\cite{1913_2016ApJ} got a theoretical prediction in GR,
$\dot{P}_{b}^{\rm GR}=-(2.40263 \pm0.00005)\times10^{-12}$.
Table~\ref{tab:PSR1913} shows the value of the observed decay rate of orbital
period $\dot{P}_{b}^{\rm obs}$, and it must be corrected by terms
$\dot{P}_{b}^{\rm Shk}$, the dynamical contributions from the Shklovskii effect
and $\dot{P}_{b}^{\rm Gal}$, the differential Galactic acceleration between the
SSB and the pulsar system.  After subtracting the total contribution from
$\dot{P}_{b}^{\rm Shk}$ and $\dot{P}_{b}^{\rm Gal}$,
$-(0.025\pm0.004)\times10^{-12}$, an intrinsic value is obtained,
$\dot{P}_{b}^{\rm intr}=-(2.398\pm0.004)\times10^{-12}$ \cite{1913_2016ApJ}.
Using $\dot{P}_{b}^{\rm GR}$ and $\dot{P}_{b}^{\rm intr}$, Weisberg and Huang
found that the ratio of the observed orbital period decrease caused by the GW
damping to its GR prediction is \cite{1913_2016ApJ},
\begin{equation}
    \frac{\dot{P}_{b}^{\rm intr}}{\dot{P}_{b}^{\rm GR}}=\frac{-(2.398\pm0.004)\times10^{-12}}{-(2.40263 \pm0.00005)\times10^{-12}}=0.9983\pm0.0016\,.
\end{equation}
This result provided the most precise test of GW emission in 2016.  One can
notice that the uncertainty of $\dot{P}_{b}^{\rm intr}$ is dominated by the
errors of $\dot{P}_{b}^{\rm Shk}$ and $\dot{P}_{b}^{\rm Gal}$, which depend on
the precision of the pulsar distance measurement and the Galactic acceleration
model.

\begin{figure}[t]
	\centering
	\includegraphics[width=7.5cm]{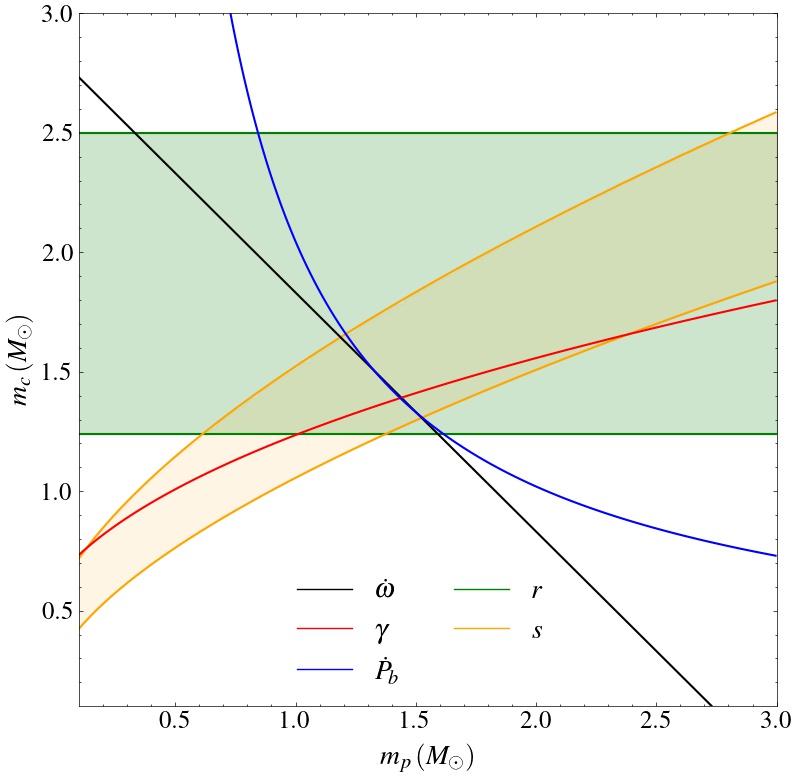}
	\caption{Mass-mass diagram of PSR~B1913+16 \cite{1913_2016ApJ}, assuming
	that GR is right and using the five PPK parameters listed in
	Table~\ref{tab:PSR1913}. The width of each band represents the $\pm
	1\mbox{-}\sigma$ uncertainty.  For $\dot{\omega}$, $\gamma$ and
	$\dot{P}_{b}^{\rm intr}$, the uncertainties of them are smaller than the
	widths of the lines. The intersection of all bands is consistent within a
	small region, which illustrates the agreement of the observations with GR.
	\label{fig:1913mpmc} }
\end{figure}

As mentioned above, assuming GR is correct and using the measured two PPK
parameters, one can get $m_{p}$ and $m_{c}$.  If one measures three or more PPK
parameters, one can do self-consistency checks in GR.  Mass-mass diagram
illustrates the self-consistency checks.  Using
Eqs.~(\ref{eq:omegadot}--\ref{eq:pbdot}), one can plot the $m_{p}$-$m_{c}$
curves for each observed PPK parameter and check whether they intersect at a
common region within measurement uncertainties.  Figure~\ref{fig:1913mpmc} shows
the mass-mass diagram of PSR~B1913+16 \cite{1913_2016ApJ}, and the curves of
five PPK parameters all meet within measurement uncertainties. For PSR~B1913+16,
three PPK parameters, $\dot{\omega}$, $\gamma$ and $\dot{P}_{b}^{\rm intr}$, are
measured with high precision; the 1-$\sigma$ errors of them are smaller than the
widths of the curves in the figure. All lines intersect in a tiny region, which
provides a strict test of GR in the strong-field condition.

\subsection{The double pulsar system: PSR~J0737$-$3039A/B}
\label{sec:doublepulsar}

PSR~J0737$-$3039A/B, discovered in 2003, is currently the only known double
pulsar system in which both NSs are detectable as radio pulsars
\cite{Burgay:2003Natur}.  PSR~J0737$-$3039A is a $22.7$-${\rm ms}$ MSP pulsar
which was first-born and spun up by accreting materials from the companion;
PSR~J0737$-$3039B is the second-born pulsar with $P=2.7\,{\rm s}$ and, due to
the geodetic precession, its radio emission disappeared in 2008
\cite{Breton2008}. The two pulsars orbit each other in a mildly eccentric
($e=0.088$) orbit with $P_{b}=2.45\,{\rm hr}$
\cite{Lyne:2004Sci,Kramer:2006Sci}.  The double pulsar system's orbital period
is only one-third of that of the Hulse-Taylor pulsar, which means
PSR~J0737$-$3039A/B has a higher average orbital velocity and a larger
acceleration.  So it is a more relativistic binary system and an excellent
testbed for strong-field gravity when well timed.

In 2006, Kramer et al.~\cite{Kramer:2006Sci} reported 2.5-yr timing results of
PSR~J0737$-$3039A/B, which provided four independent strong-field tests of GR.
Although only using 2.5 years of timing data, because of the larger relativistic
effect in this system, the PPK parameters $\dot{\omega}$, $\gamma$ and
$\dot{P}_{b}$ were already well measured.  The orbit is nearly edge-on, so the
PPK parameters $r$ and $s$ of the Shapiro delay effect also have precise
measurements.  Because the two pulsars' emissions are detectable for this
system, the projected semi-major orbital axes $x_{\rm A}$ and $x_{\rm B}$ are
both measured with high precision. Therefore this system provides an accurate
measurement of the mass ratio $R=m_{\rm A}/m_{\rm B}=x_{\rm B}/x_{\rm A}$, which
is a theory-independent parameter in boost-invariant gravity theories.  Kramer
et al.~\cite{Kramer:2006Sci} used the most precisely measured PPK parameter
$\dot{\omega}=16.89947(68)\,{\rm deg\,yr^{-1}}$ and the theory-independent
parameter $R=1.0714(11)$ to derive the masses of the binary, $m_{\rm
A}=1.3381(7)\,M_{\odot}$ and $m_{\rm B}=1.2489(7)\,M_{\odot}$. The masses are
then used to calculate the theoretical values of other PPK parameters with the
assumption that GR is right.  In Table~\ref{tab:J0737-06} we show the results of
four independent gravity tests.  The fourth column is the ratio of observed
values to their expectation in GR, wherein the ratio for $s$ was the best
available Shapiro-delay test of GR in the strong-field region at that time. With
only 2.5-yr data, it reveals PSR~J0737$-$3039A/B's outstanding ability to test
GR.

\begingroup
\setlength{\tabcolsep}{10pt}
\renewcommand{\arraystretch}{1.2}
\begin{table}[t]
\begin{center}
\caption{Four independent tests of GR provided by the PPK parameters of the
double pulsar \cite{Kramer:2006Sci}.  The last column shows the ratios of
observed values to their expectation in GR.}
\label{tab:J0737-06}       
\begin{tabular}{p{2cm}p{2cm}p{2cm}p{1.8cm}}
\noalign{\smallskip}\svhline\noalign{\smallskip}
PPK Parameter & Observed &  Expected in GR & Ratio  \\
\hline\noalign{\smallskip}
$\dot{P}_{b}$ $(10^{-12})$  &$-1.252(17)$ & $-1.24787(13)$ & 1.003(14) \\
$\gamma ~{\rm (ms)}$ & 0.3856(26) & 0.38418(22) & 1.0036(68) \\
$s$ & $0.99974(-39,\,+16)$ & $0.99987(-48,\,+13)$ & 0.99987(50) \\
$r$ ($\mu$s) & 6.21(33) & 6.153(26) & 1.009(55) \\
\noalign{\smallskip}\hline\noalign{\smallskip}
\end{tabular}
\end{center}
\end{table}
\endgroup

For PSR~J0737$-$3039A/B, it shows a roughly 30-s eclipse when pulsar A passes
behind pulsar B, namely, pulsar A is at the superior conjunction of its orbit.
Pulsar A's pulse signals will be absorbed by the magnetosphere of pulsar B when
the eclipse happens.  For pulsar B, the misalignment angle between the spin
vector and the total angular momentum vector is significant,
$\delta_{B}\simeq50^{\circ}$, leading to a measurable geodetic precession of
pulsar B's spin, $\textbf{S}_{B}$.  The geodetic precession of $\textbf{S}_{B}$
causes a changing absorption effect at the eclipse. Breton et
al.~\cite{Breton2008} used a simple geometric model to characterize the observed
changing eclipse morphology and measured the geodetic precession of
$\textbf{S}_{B}$ around the total orbital angular momentum.  They got a
relativistic spin precession rate,
$\Omega_{B}=4.77^{\circ}(+0.66^\circ,\,-0.65^\circ)\,{\rm yr^{-1}}$, and this
result is consistent with the value, $\Omega_{B}^{\rm
GR}=5.0734(7)^{\circ}\,{\rm yr^{-1}}$, predicted by GR, providing a new test of
GR in the strong-field regime \cite{Breton2008}.

In 2021, Kramer et al.~\cite{Kramer:2021jcw} reported 16.2-yr timing data of the
double pulsar observed with six radio telescopes. These results show an
excellent ability to test gravity theories because they not only provide very good test of GR in the strong field but also reveal new relativistic effects
that are observed for the first time.  These timing results provide seven PPK
parameters, which are more than those from any other known binary pulsars, and
some of the parameters need to take higher-order contributions into account. Our
following discussion will highlight some novel aspects of the double pulsar from
this new set of data.

As mentioned before, the DD model is a compact timing model that uses the
solution of the 1\,PN two-body problem. However, the precision of timing for the
double pulsar is so high that higher-order corrections to the timing model need
to be added.  For $\dot{\omega}$, 2\,PN and the Lense-Thirring effects are
needed~\cite{Hu:2020ubl}
\begin{align}
    \dot{\omega} = \dot{\omega}_{\rm 1\,PN} + \dot{\omega}_{\rm 2\,PN} + \dot{\omega}_{{\rm LT}_{\rm A}}\,,
\end{align}
where $\dot{\omega}_{\rm 2\, PN} =\dot{\omega}_{\rm 1\,PN}f_{\rm O}
{V_{b}^{2}}/{c^{2}}$ with
\begin{align}
 f_{\rm O} &= \frac{1}{1-e^{2}}\left(\frac{3}{2}X_{A}^{2}+\frac{3}{2}X_{A}+\frac{27}{4}\right)+\left(\frac{5}{6}X_{A}^{2}-\frac{23}{6}X_{A}-\frac{1}{4}\right)\,,
\end{align}
where $X_{\rm A}\equiv m_{\rm A}/m$.  For PSR~J0737$-$3039A, the misalignment
angle of $\textbf{S}_{\rm A}$ relative to the total angular momentum vector is
very small, $\delta_{\rm A}<3.2^{\circ}$~\cite{Ferdman:2013xia}.  So
Eq.~(\ref{eq:omdotLT}) is used to calculate $\dot{\omega}_{{\rm LT}_{\rm A}}$.
Using values from timing, it was found that $\dot{\omega}_{\rm
2\,PN}\simeq4.39\times10^{-4}\,{\rm deg\,yr^{-1}}$ in GR, about 35 times of the
measurement error of $\dot{\omega}$ (see Table~\ref{tab:PSRJ0737}).  For
$\dot{\omega}_{{\rm LT}_{\rm A}}$, except that the moment of inertia of pulsar A
($I_{\rm A}$)  depends on the EOS of NSs, all quantities in
Eq.~(\ref{eq:omdotLT}) are measured with high precision, and we get
\cite{Kramer:2021jcw},
\begin{equation}
    \dot{\omega}_{\mathrm{LT}_\mathrm{A}} \simeq-3.77 \times 10^{-4} \times I_{\mathrm{A}}^{(45)}\,\mathrm{deg}\,\mathrm{yr}^{-1}\,,
\end{equation}
where $I_{\mathrm{A}}^{(45)} \equiv I_{\mathrm{A}}/(10^{45}\,{\rm g\,cm^{2}})$.
Generally, $I_{\mathrm{A}}^{(45)}$ is of order unit, so $\dot{\omega}_{{\rm
LT}_{\rm A}}$ and $\dot{\omega}_{\rm 2\,PN}$ have the same order of
magnitude~\cite{Kramer:2009zza}.

\begingroup
\setlength{\tabcolsep}{10pt}
\renewcommand{\arraystretch}{1.2}
\begin{table}[t]
\begin{center}
\caption{Orbital parameters of PSR J0737$-$3039A \cite{Kramer:2021jcw}.}
\label{tab:PSRJ0737}       
\begin{tabular}{p{5.5cm}p{3cm}}
\noalign{\smallskip}\svhline\noalign{\smallskip}
Parameter & Value  \\
\hline\noalign{\smallskip}
Projected semi-major axis, $x\,({\rm s})$ & 1.415028603(92) \\
Eccentricity, $e$ & 0.087777023(61) \\
Orbital period, $P_{b}\,({\rm d})$ & 0.1022515592973(10)\\
Epoch of periastron, $T_{0}\,({\rm MJD})$ & 55700.233017540(13) \\
Longitude of periastron, $\omega_0 \, ({\rm deg})$ & 204.753686(47) \\
Periastron advance rate, $\dot{\omega}\,({\rm deg\,yr^{-1}})$ & 16.899323(13) \\
Einstein delay amplitude, $\gamma\,({\rm ms})$ & 0.384045(94) \\
Change rate of orbital period, $\dot{P}_{b}$ & $ -1.247920(78) \times 10^{-12}$ \\
Logarithmic Shapiro shape, $z_{s}$ & 9.65(15)\\
Range of Shapiro delay, $r\,({\rm \mu s})$ & 6.162(21) \\
NLO factor for signal propagation, $q_{\rm NLO}$ & 1.15(13)\\
Relativistic deformation of orbit, $\delta_{\theta}$ & $ 13(13) \times 10^{-6}$ \\
Change rate of projected semi-major axis, $\dot{x}$ & $8(7)\times10^{-16}$ \\
Change rate of eccentricity, $\dot{e}\,({\rm s}^{-1})$ & $3(6)\times10^{-16}$ \\
\noalign{\smallskip}\hline\noalign{\smallskip}
Derived parameters & ~\\
\noalign{\smallskip}\hline\noalign{\smallskip}
$\sin{i}=1-{\rm exp}(-z_{s})$ & $0.999936(+9,\,-10)$\\
Orbital inclination, $i\,({\rm deg})$  & 89.35(5) or 90.65(5)\\
Total mass, $m\,(M_{\odot})$ & $2.587052(+9,\,-7)$ \\
Mass of pulsar A, $m_{\rm A}\,(M_{\odot})$ & $1.338185(+12,\,-14)$\\
Mass of pulsar B, $m_{\rm B}\,(M_{\odot})$ & $1.248868(+13,\,-11)$\\
\noalign{\smallskip}\hline\noalign{\smallskip}
\end{tabular}
\end{center}
\end{table}
\endgroup

In Eq.~(\ref{eq:romer}), the R$\ddot{\rm o}$mer delay depends on the projected
semi-major axis $x$.  In Newtonian gravity, the binary mass function describes
the relation between the inclination angle $i$ and $x$ by Eq.\,(\ref{eq:s}).
Because of the high precision measurements, the mass function now needs to
extend to 1\,PN \cite{Kramer:2021jcw},
\begin{equation}\label{eq:massfun1pn}
    \sin i=\frac{n_{\mathrm{b}} x}{X_{\mathrm{B}}}\frac{c}{V_{b}}
    \left[1+\left(3-\frac{1}{3} X_{\mathrm{A}} X_{\mathrm{B}}\right) \frac{V_{b}^{2}}{c^{2}}\right]\,,
\end{equation}
with $X_{\rm B}=m_{\rm B}/m$. Such an extension was the first time for any binary pulsar system.

GW damping enters the GR equations of motion at 2.5\,PN level [c.f.\
Eq.~(\ref{eq:pbdot})]. For the double pulsar, it is extended to 3.5\,PN, with
$\dot{P}_{b}^{3.5\,{\rm PN}}=\dot{P}_{b}^{2.5\,{\rm PN}}X^{3.5\,{\rm PN}} \simeq
-1.75\times10^{-17}$; for the factor $X^{3.5\,{\rm PN}}$ one can refer to
Eqs.~(12--13) in Ref.~\cite{Hu:2020ubl}.  This value is about 4.5 times smaller
than the timing precision of $\dot{P}_{b}$.  The mass loss caused by the
spin-down of the pulsars can also contribute to the observed $\dot{P_{b}}$, and
for pulsar A, it is $\dot{P}_{\mathrm{b}}^{\dot{m}_{\mathrm{A}}}=2.3 \times
10^{-17} \times I_{\mathrm{A}}^{(45)}$.  Using constraints from
Ref.~\cite{Dietrich:2020Sci} and the radius-$I_A$ relation in
Ref.~\cite{Lattimer:2019Univ}, one gets
$\dot{P}_{\mathrm{b}}^{\dot{m}_{\mathrm{A}}}=2.9(2) \times 10^{-17}$, which is 3
times smaller than the error of $\dot{P}_{b}^{\rm int}$.  Besides
$\dot{P}_{b}^{\rm int}$, the non-intrinsic contributions
$\dot{P}_{\mathrm{b}}^{\mathrm{Shk}}$ and $\dot{P}_{\mathrm{b}}^{\mathrm{Gal}}$
also need to be considered.  These two  terms are both functions of the pulsar
distance $d$. The parallax from VLBI and pulsar timing was combined to provide a
weighted distance, $d=(735\pm60)\,{\rm pc}$. Considering all the above,
$\dot{P_{b}}$ is expressed as, 
\begin{equation}
	\dot{P_{b}} = \dot{P}_{b}^{2.5\,{\rm PN}} + \dot{P}_{b}^{3.5\,{\rm PN}} +
	\dot{P}_{\mathrm{b}}^{\dot{m}_{\mathrm{A}}}
	+\dot{P}_{\mathrm{b}}^{\mathrm{ext}}  \,,
\end{equation}
where
$\dot{P}_{\mathrm{b}}^{\mathrm{ext}} = \dot{P}_{\mathrm{b}}^{\mathrm{Shk}} +
\dot{P}_{\mathrm{b}}^{\mathrm{Gal}}=-1.68^{+11}_{-10} \times 10^{-16}$, leaving
the intrinsic contribution, $\dot{P}_{\mathrm{b}}^{\mathrm{int}}= \dot{P_{b}} -
\dot{P}_{\mathrm{b}}^{\mathrm{ext}} =
-1.247752(79)\times10^{-12}$~\cite{Kramer:2021jcw}.  The error of
$\dot{P}_{\mathrm{b}}^{\mathrm{int}}$ is still dominated by the error of
$\dot{P_{b}}$. In the future improving the timing precision of the observed
$\dot{P_{b}}$ will improve this gravity test.

The Shapiro delay effect in Eq.~(\ref{eq:Shapiro}) is the leading-order effect
caused by the companion's mass influencing the  signal propagation.  It was
obtained by integrating along a straight line and assuming a static mass
distribution~\cite{Blandford:1976}.  For the double pulsar one needs a lensing
correction to the Shapiro delay. It restores the fact that pulsar A's radio
signal propagates along a curved path.  Equation~(\ref{eq:Shapiro}) is rewritten
as ${\rm \Delta}_{\rm SB}=-2r\ln{\Lambda_{u}}$, and Kramer et
al.~\cite{Kramer:2021jcw} added a term corresponding to the lensing correction
$\delta\Lambda_{u}^{\rm len}=2rc/a_{R}$, where $a_{R}$ is the semi-major axis of
the relative orbit.  Considering a non-static mass distribution,  a 1.5\,PN
correction $\delta\Lambda_{u}^{\rm ret}$ from the retardation effect was taken
into account.  So the signal propagation delay is extended to ${\rm
\Delta}_{\rm SB} = -2r\ln\left(\Lambda_{u}+\delta\Lambda_{u}^{\rm
len}+\delta\Lambda_{u}^{\rm ret}\right)$.  For the aberration delay ${\rm
\Delta}_{\rm AB} $, considering the rotational deflection delay, Kramer et
al.~\cite{Kramer:2021jcw} used a lensing correction term, $\mathcal{D}
{\cos{\Phi}}/{\Lambda_{u}}$, where $ \Phi = \omega+A_{e}(u)$.  Nevertheless, the
next-to-leading-order (NLO) contributions from the Shapiro delay and the
aberration delay cannot be tested separately in the double pulsar system.
These contributions are rescaled with a common factor $q_{\rm NLO}$ to test the
significance of NLO signal propagation contributions.  The factor $q_{\rm NLO}$
can be fitted in the extended timing model with the double pulsar data.

After providing an extended timing model with higher-order contributions, one
can use the timing results in Table~\ref{tab:PSRJ0737} to determine the masses
and test GR.  Generally, one needs two PPK parameters to determine the  masses of
a binary system in GR.  However, when including higher-order contributions, the
parameters $\dot{\omega}$ and $\dot{P}_{b}^{\rm int}$ are also functions of
$I_{A}$, namely $\dot{\omega}=\dot{\omega}(m_{A},m_{B},I_{A})$ and
$\dot{P}_{b}^{\rm int}=\dot{P}_{b}^{\rm int}(m_{A},m_{B},I_{A})$.  Using
constraints on the EOS obtained from the binary NS merger event GW170817
\cite{LIGOScientific:2017vwq}, one has restriction on $I_{A}$.  Kramer et
al.~\cite{Kramer:2021jcw} used the information of $I_{A}$ and the timing values
of $\dot{\omega}$ and $s$ and got the masses of the binary,
\begin{align}
     m_{\mathrm{A}}=\,&1.338185(+12,\,-14)\,M_{\odot}\,,\label{eq:ma} \\
     m_{\mathrm{B}}=\,&1.248868(+13,\,-11)\,M_{\odot}\,, \label{eq:mb}\\
     m =\,&2.587052(+9,\,-7)\,M_{\odot}\,.
\end{align}

If one ignores any existing constraints on the EOS of NSs and simultaneously
determine $m_{A}$, $m_{B}$ and $I_{A}$, one needs three PPK parameters.  With
$\dot{\omega}(m_{A},m_{B},I_{A})$, $\dot{P}_{b}^{\rm int}(m_{A},m_{B},I_{A})$
and $s(m_{A},m_{B})$, one not only gets the masses of the system, but also
constrains $I_A$ to $I_{A}<3.0\times10^{45}\,{\rm g\,cm^{2}}$ with $90\%$
confidence, complementing the constraints from
LIGO/Virgo~\cite{LIGOScientific:2017vwq,Landry:2018ApJ} and
NICER~\cite{Silva:2020acr} observations.

\begin{figure}[t]
	\centering
	\includegraphics[width=11.5cm]{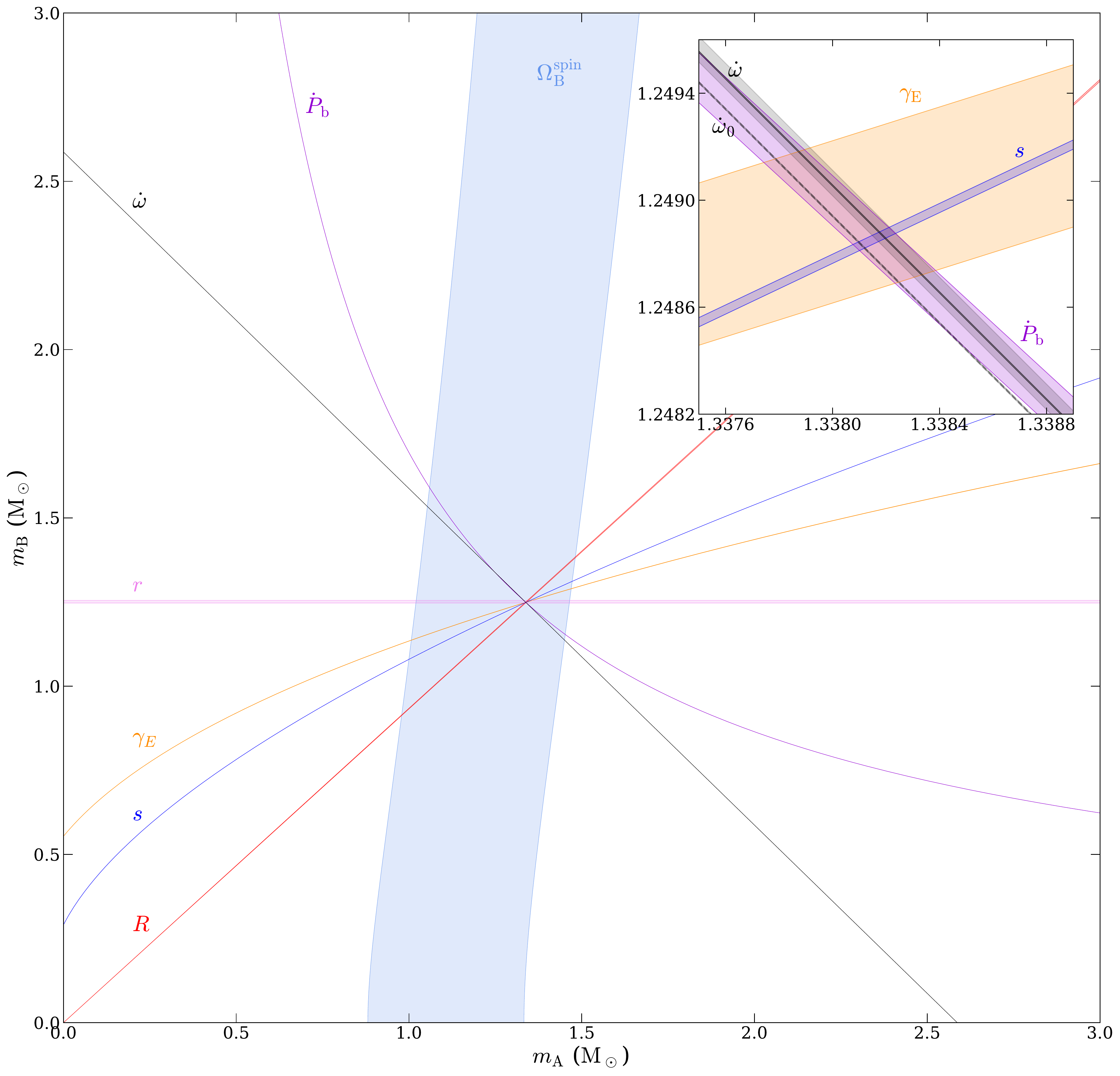}
	\caption{Mass-mass diagram of PSR~J0737$-$3039 \cite{Kramer:2021jcw}. It
	shows the constraints from six PPK parameters and the mass ratio $R$.  The
	inset shows the zoom-in region of the intersection of these measurements.
	The solid black line for $\dot{\omega}$ responds to $I_{A}^{(45)}=1.32$. The
	gray band indicates the range for $\dot{\omega}$ with a changing
	$I_{A}^{(45)}$. The dashed black line for $\dot{\omega}_{0}$ does not
	contain the contribution of the Lense-Thirring effect. \label{fig:0737mpmc}
	}
\end{figure}

The double pulsar provides seven PPK parameters,
$\{\dot{\omega},\,\gamma,\,\dot{P}_{b},\,r,\,s,\,\Omega_{B},\,\delta_{\theta}\}$,
and a theory-independent parameter $R$ (c.f.\ Table~\ref{tab:PSRJ0737}).  For
$\delta_{\theta}$, its measured value is not significant, so it is not used in
testing GR.  When a value for $I_{A}$ is chosen, these PPK parameters are
functions of binary masses and those well-measured Keplerian parameters.  So
using six measured PPK parameters and the mass ratio $R$, one can perform five
independent tests of GR, as shown in Fig.~\ref{fig:0737mpmc}
\cite{Kramer:2022gru}.

As mentioned earlier, including the 3.5\,PN contribution from GW damping and
removing the mass loss and non-intrinsic contributions, one has
\cite{Kramer:2021jcw},
\begin{equation}
    \dot{P}_{\mathrm{b}}^{\mathrm{GW}}=\dot{P}_{\mathrm{b}}^{\mathrm{int}}-\dot{P}_{\mathrm{b}}^{\dot{m}_{\mathrm{A}}}=-1.247782(79) \times 10^{-12}\,.
\end{equation}
Using the masses shown in Eqs.~(\ref{eq:ma}--\ref{eq:mb}), the predicted value
from GR is,
\begin{align}
\dot{P}_{\mathrm{b}}^{\mathrm{GW}, \mathrm{GR}} & =\dot{P}_{\mathrm{b}}^{\mathrm{GW}, \mathrm{GR}(2.5 \, \mathrm{PN})}+\dot{P}_{\mathrm{b}}^{\mathrm{GW}, \mathrm{GR}(3.5 \, \mathrm{PN})}  =-1.247827(+6,\,-7) \times 10^{-12}\,.\nonumber 
\end{align}
This value provides the most precise test for the GW quadrupolar emission
\cite{Kramer:2021jcw}, $\dot{P}_{b}^{\rm GW}/\dot{P}_{b}^{\rm
GW,GR}=0.999963(63)$.  It is also the most precise result among the different
independent tests that one can obtain from this system. It is in agreement with
GR at a level of $1.3\times10^{-4}$ with $95\%$ confidence
\cite{Kramer:2021jcw}.  Other independent tests are listed in
Table~\ref{tab:PSRJ0737GRtest}.  PSR~J0737$-$3039A/B provides the strictest
limit on GR than any other binary pulsar systems.

\begingroup
\setlength{\tabcolsep}{10pt}
\renewcommand{\arraystretch}{1.2}
\begin{table}[t]
\begin{center}
\caption{Relativistic effects measured in the double pulsar system and the
resulting independent strong-field tests of GR \cite{Kramer:2021jcw}. The right
column shows the ratio between the observed quantity and its prediction in GR.}
\label{tab:PSRJ0737GRtest}       
\begin{tabular}{p{5.5cm}p{3cm}}
\noalign{\smallskip}\svhline\noalign{\smallskip}
Parameter & Ratio  \\
\hline\noalign{\smallskip}
Shapiro delay shape, $s$ & 1.00009(18) \\
Shapiro delay range, $r$ & 1.0016(34) \\
Time dilation, $\gamma$ & 1.00012(25)\\
Periastron advance, $\dot{\omega}$ & 1.000015(26) \\
Change rate of orbital period, $\dot{P}_{b}$ & 0.999963(63) \\
Orbital deformation, $\delta_{\theta}$ & 1.3(13)\\
Spin precession, $\Omega_{\rm B}$ & 0.94(13) \\
\noalign{\smallskip}\hline\noalign{\smallskip}
Tests of higher-order contributions &~\\
\noalign{\smallskip}\hline\noalign{\smallskip}
Lense-Thirring contribution to $\dot{\omega}$, $\lambda_{\rm LT}$ & 0.7(9)\\
NLO signal propagation, $q_{\rm NLO}$ [total] & 1.15(13)\\
\noalign{\smallskip}\hline\noalign{\smallskip}
\end{tabular}
\end{center}
\end{table}
\endgroup

\subsection{The triple system: PSR~J0337+1715}
\label{sec:triple}

Binary pulsar systems provide us with great opportunities to test GR.  There are
also some pulsars that exist in more complicated systems.  For example,
PSR~B1257+12, a pulsar moves in a system that contains at least three low-mass
planets.  This pulsar helped people discover the first exoplanet system
\cite{Wolszczan:1992zg, Malhotra:1992}. Here we focus on PSR~J0337+1715, which
is a $2.7$-${\rm ms}$ pulsar in a hierarchical triple system with two white
dwarf (WD) companions \cite{Ransom:2014}.  The inner pulsar-WD system is a 1.6-d
circular orbit system and the mass of the WD is $0.19751(15)\,M_{\odot}$.  The
outer orbit is wider, with $P_{b}=327.2\,$d, and the WD's mass is
$0.4101(3)\,M_{\odot}$ \cite{Ransom:2014}.  The special system composition and
high-precision timing results of PSR~J0337+1715 make it an ideal laboratory that
can provide a strong equivalence principle (EP) test of GR
\cite{Archibald:2018oxs,Voisin:2020lqi}.

The principle of equivalence has played an important role in the development of
theories of gravity.  In Newton's gravity, the principle of equivalence is the
cornerstone of the theory. It assumes that the inertial mass equals the passive
gravitational mass.  Now, this is so-called the weak EP (WEP), and its
alternative statement is that the trajectory of a freely falling test body is
independent of the body's internal structure and composition.  With the
development of gravity theory, the principle of equivalence has been extended.
Except for satisfying the WEP, one believes that a gravity theory should also
satisfy the local Lorentz invariance (LLI) and local position invariance (LPI).
Combined with WEP, we call it the Einstein EP (EEP).  In the EEP, LLI states
that a non-gravitational experiment does not depend on the velocity of the
freely falling reference frame, and LPI states that the outcome of any local
non-gravitational experiment is independent of when and where the experiment is
performed.  Metric theories of gravity satisfy the EEP~\cite{Will:2018bme}.  The
strictest bound comes from the strong EP (SEP), it demands that WEP is valid for
self-gravitating bodies as well, and a gravitational experiment needs to embody
LLI and LPI simultaneously.  For now, GR seems to be the only viable metric
theory that satisfies SEP completely. So testing SEP is equal to testing GR in
this sense.

There are many methods to test SEP, and one of them is considering the
universality of free fall of two self-gravitating bodies in an external
gravitational field.  Some tests from the Solar System have been done and
provided tight constraints on the violation of the SEP, but the bodies in the
Solar System are weak in terms of self-gravity.  For instance, the fractional
binding energy difference between the Earth and the Moon is $\sim-4 \times
10^{-10}$.  Pulsar-WD systems can provide an evident fractional binding energy
difference between the NS and the WD, at the order of $-0.1$, which is many
orders of magnitude larger than the Earth-Moon system's difference.  If the
acceleration difference between the pulsar and the WD is present, it will lead
to a characteristic change in its orbital eccentricity evolution.  However, a
pulsar-WD system's free fall in the Galactic gravitational field cannot provide
a tight limit on the SEP violation parameter $\Delta$, because the Galactic
gravitational acceleration at the location of the pulsar is comparably weak,
$\sim 2\times10^{-8} \, {\rm cm\,s^{-2}}$. For example, PSR~J1713+0747 provides
a constraint, $\Delta<0.002$ \cite{Zhu:2018etc}.  For the triple system
PSR~J0337+1715, the outer WD can provide an external gravitational acceleration
for the inner PSR-WD system, and its strength is $\sim 0.17\,{\rm cm\,s^{-2}}$,
which is $\sim10^{7}$ times larger than the Galactic gravitational acceleration.
So PSR~J0337+1715 makes a significant improvement on the constraint of the SEP
violation parameter, $\Delta<2.6\times10^{-6}$ \cite{Archibald:2018oxs} and
$\Delta = (0.5 \pm 0.9)\times 10^{-6}$~\cite{Voisin:2020lqi} by two independent
groups.

\section{Neutron Star Structures}
\label{sec:ns}

As introduced in the previous section, pulsars, which are rotating NSs, can
provide unique possibilities to probe the strong-field region of GR. With a
typical mass of $1.4\ \rm{M_\odot}$ and a radius of 12 km, NSs not only provide
the strong-field environment of gravity, but also are the most extreme
laboratory of particle physics. In this section, we will introduce the basic
theory describing the NS structures, and show some properties of a single NS
that can be measured with pulsar timing technique and other observations. 
Differing from the tests of gravity from binary pulsar systems, which use the
dynamic aspect of the gravity theory, testing gravity with a single NS provides
the properties of an equilibrium state.  In the following, we take $G=c=1$
except when the units are written out explicitly, and the convention of the
metric is $(-,+,+,+)$.

The basic information of a compact star is its mass and radius, which reveal
the inner structure and interaction among the star's components. In GR, to
derive the structure of a spherical, non-rotating NS, we begin with the
Einstein's field equations, $G_{\mu\nu}=8\pi T_{\mu\nu}$, together with some
specific matter model, for example a perfect fluid,
$T^{\mu\nu}=(\rho+p)u^{\mu}u^{\nu}+pg^{\mu\nu}$.  Then one adopts the metric of
a static, spherically symmetric spacetime
\begin{eqnarray}
	{\rm d}s^2=-\left(1-\frac{2m(r)}{r}\right){\rm d}t^2+\frac{{\rm d}r^2}{1-2m/r}+r^2{\rm d}{\rm\Omega}^2\,,
\end{eqnarray}
and inserts it into the field equations. 
The equation of a star is described by the Tolman-Oppenheimer-Volkov (TOV)
equation~\cite{Oppenheimer:1939ne,Tolman:1939jz}

\begin{eqnarray}\label{eq: TOV}
	\frac{{\rm d}p}{{\rm d}r}&=&-\frac{(\rho+p)(m+4\pi r^3 p)}{r(r-2m)}\,,
\end{eqnarray}
where $m(r)$ is called the mass inside the sphere of radius $r$, defined by the
integral, $m(r)=\int_0^r 4\pi r^2\rho {\rm d}r$, just as in the Newtonian
theory.  Results show that for any physical realizable matter, $m(r)/r\leq
0.485$, so that the right hand side of Eq.~(\ref{eq: TOV}) is never
singular~\cite{Bondi:1964zz}.  The dimensionless ratio between a star's mass and
radius, $C\equiv M/R$, is called the compactness of the star, which represents
how compact a star is. For Schwarzschild black holes (BHs) in GR this number is
equal to 0.5 and for a typical NS, the compactness can reach 0.2. As a
comparison, the compactness of the Sun is only $2\times 10^{-6}$. This is
crucial for the SEP in Sec.~\ref{sec:triple}.

To complete the above equations, a relation called the EOS between the pressure
$p$ and the energy density $\rho$ is needed. For a simple fluid in local
thermodynamic equilibrium, there always exists a relation of the form,
$p=p(\rho,S)$, where $S$ is the specific entropy. For old NSs that are
sufficiently cold or under the assumption of constant entropy, we can restrict
this relation to a one-parameter functional, $p=p(\rho)$.

For a given EOS, one can then solve the TOV equation with initial condition
$\rho|_{r=0}=\rho_0$ to get the structure of a NS. With the uniqueness theorems
for ordinary differential equations, the solutions will form a one-parameter
sequence with the parameter being the central density. Correspondingly, for a
given theory of gravity, this sequence, the so-called mass-radius relation, also
uniquely determines the underlying EOS~\cite{Lindblom:1992}. The total mass of
the NS is $m(R)$, where $R$ is the radius of the surface defined by $p(R)=0$.
For a normal NS that is gravitationally self-bounded, the density also
approaches zero at the star's surface, i.e. $\rho(R)=0$.  But if a NS
consists of quarks or strangeons that are not gravitationally self-bounded,
there can be a density discontinuity at the surface of the
star~\cite{Damour:2009vw, Reina:2015jia,Gao:2021uus}.

\begin{figure}[t]
	\centering
	\subfloat[EOSs]{\label{fig: EoS}
	\includegraphics[width=8cm]{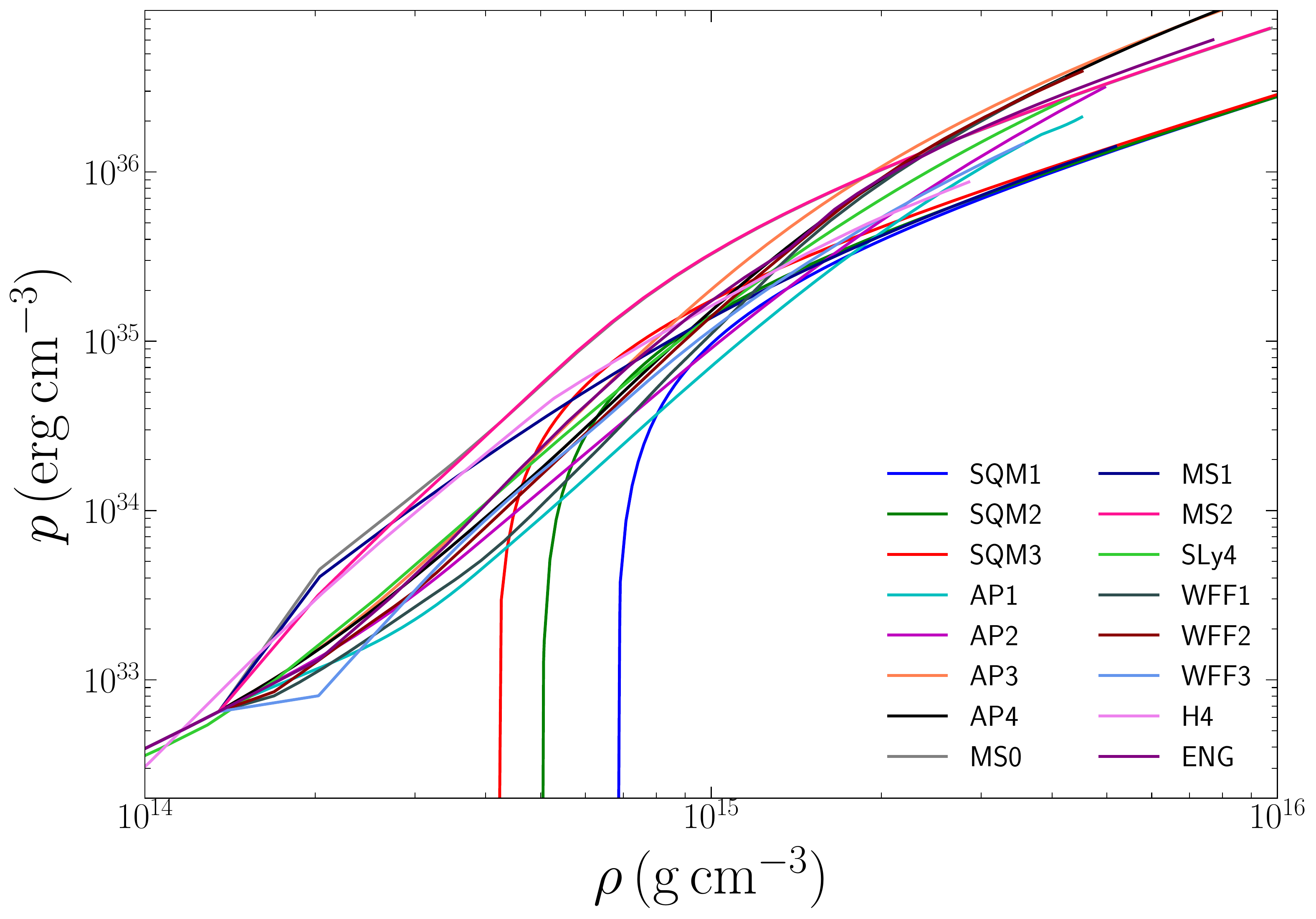}
	} \linebreak
	\subfloat[$M$-$R$ relations]{\label{fig: M-R}
	\includegraphics[width=8cm]{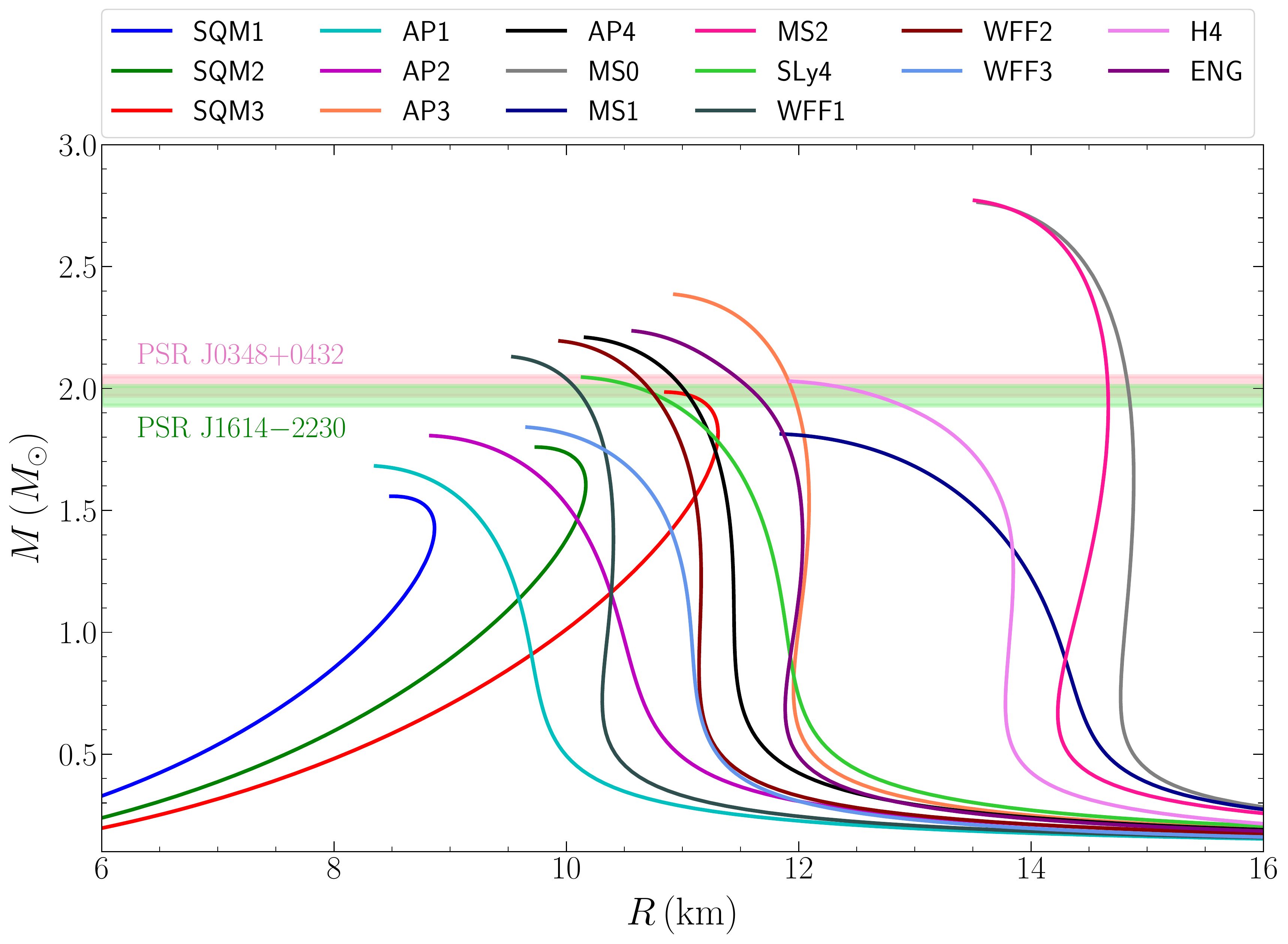}
	}
	\caption{\label{fig: EoS_M-R}EOSs for NSs~\cite{Lattimer:2000nx} and the
	related $M$-$R$ relations (figure courtesy of Y. Gao). The horizontal lines
	in Fig.~\ref{fig: M-R} are the observed masses for two massive
	pulsars~\cite{Antoniadis:2013pzd, Fonseca:2021wxt}. The SQM model is related
	to quark stars~\cite{Lattimer:2000nx}. Curves passing the highest point in
	their $M$-$R$ relations are cut off, as they represent unstable NSs. }
\end{figure}

In Fig.~\ref{fig: EoS_M-R} we show some EOSs and the corresponding $M$-$R$
relations for NSs. The EOSs inside the NSs are still not clear today. The main
components of NSs are mainly regarded as neutrons, but there also should be
protons, electrons, muons, and so on. It remains an unsolved problem in quantum
chromodynamics for calculating the super-dense nuclear matter at low
temperatures. Furthermore, there are some arguments that the ingredients of NSs
should be quarks or strangeons~\cite{Witten:1984rs, Xu:2003xe, Lai:2017ney,
Gao:2021uus}. From Fig.~\ref{fig: EoS_M-R}, we can see that there is large
difference between different EOS models, especially between the models for a
normal NS and models for a quark or a strangeon star.  For those NSs that are
not gravitationally self-bounded, the behavior at the low mass limit is like a
star with constant density. This feature still holds approximately when the mass
of the star increases, and leads to various differences in other aspects like
the moment of inertia and the star oscillation frequency (see e.g.\
Ref.~\cite{Li:2022qql}). But one can notice that all of these EOS models have a
maximum mass in their $M$-$R$ relations. For WDs this maximum mass is famously
known as the Chandrasekhar limit, which is around $1.4\,{\rm M_{\odot}}$. Due to
the uncertainty in the EOS, the maximum mass of NSs can vary largely, from about
$1.5\,{\rm M_\odot}$ to over $3\,{\rm M_\odot}$. Known as the turning point
theory~\cite{Hadzic:2020smr}, a NS with mass and radius crossed the maximum mass
point in the $M$-$R$ relation becomes unstable.  So the discover of supermassive
NSs can help us constrain the EOS of NSs. The two horizontal lines in
Fig.~\ref{fig: M-R} are the observed masses of
PSR~J0348+0432~\cite{Antoniadis:2013pzd} and
PSR~J1614$-$2203~\cite{Fonseca:2021wxt}, and then EOSs which cannot support a
maximum mass higher than $2\,{\rm M_\odot}$ are excluded by these observations. 
But if one considers gravity theories different from GR, the $M$-$R$ relation of
the same EOS can vary largely from their GR counterpart. Thus a precise
measurement of the $M$-$R$ relation can constrain not only the EOS but also the
gravity theory~\cite{Shao:2022koz}.  Previous measurements from the NICER
satellite by modeling the thermal x-ray waveform of the isolate pulsar
PSR~J0030+0451 give a constraint on the mass and equatorial radius of the star
simultaneously~\cite{Riley:2019yda,Miller:2019cac}. But measurements from more NSs with higher
precision are still needed.

The above discussion is based on the assumption of a static star. For real
pulsars, they can have spin periods from several milliseconds to tens of seconds
(see Fig.~\ref{fig:ppdot}).  Some studies have shown that fast-rotating NSs can
support a larger maximum mass and it will have important influence on the
evolution of the binary NS merger remnant~\cite{Bauswein:2017aur,Zhou:2021tgo}.
One can estimate the maximum angular velocity of a NS from the Newtonian theory,
$\Omega_{\rm max}=(M/R^3)^{1/2}$, which is related to the mass shedding limit.
For a typical NS, this corresponds to a spin period of about 0.5\,ms. Thus for
most pulsars, the approximation of slow rotation is sufficient, and in this
case, the mass-radius relation of NSs will not change, to the first order in
$\Omega$.

The basic property related to rotation is the moment of inertia. In GR, the
moment of inertia of a slowly rotating NS can be calculated through the slow
rotation approximation~\cite{Hartle:1967he}. Under this approximation the metric
is writen as
\begin{eqnarray}
	{\rm d}s^2=g_{\mu\nu}{\rm d}x^{\mu}{\rm d}x^{\nu}&=&-\left(1-\frac{2m}{r}\right){\rm d}t^2+\frac{{\rm d}r^2}{1-2m/r}+r^2\left({\rm d}\theta^2+\sin^2\theta {\rm d}\phi^2\right)\nonumber\\
	&&+r^2\sin^2\theta\left({\rm d}\phi+\left(\omega(r)-\Omega\right){\rm d}t\right)^2\,,
\end{eqnarray}
where $m(r)$ is just the same as that solved from the TOV equation for a
non-rotating NS, $\Omega$ is the angular velocity of the star and $\omega(r)$
shows the effect of frame dragging. An observer falling freely from infinity to
$r$ will have an angular velocity $\Omega-\omega(r)$ to the first order in
$\Omega$, thus seeing the fluid element at $r$ has an angular velocity
$\omega(r)$.  Function $\omega(r)$ is determined by the linear differential
equation,
\begin{eqnarray}\label{eq: omega}
	\frac{{\rm d}^2\omega}{{\rm d}r^2}&=&\frac{4\pi r}{r-2m}(\rho+p)\left(r\frac{{\rm d}\omega}{{\rm d}r}+4\omega\right)-\frac{4}{r}\frac{{\rm d}\omega}{{\rm d}r}\,.
\end{eqnarray}
The boundary condition is $\omega=\Omega$ when $r$ goes to infinity. Besides
that, in order to keep ${\rm d}^2\omega/{\rm d}r^2$ to be finite at the origin,
we have ${\rm d}\omega/{{\rm d}r}|_{r=0}=0$. The angular momentum $J$ carried by the star can be read
out from the asymptotic behavior of $g_{t\phi}$ at $r\rightarrow\infty$,
	$g_{t\phi}=(\omega-\Omega)r^2\sin^2\theta\rightarrow-{2J\sin^2\theta}/{r}$,
which indicates that
\begin{equation}
	J=\frac{1}{6}\frac{{\rm d}\omega}{{\rm d}r}r^4 \Big|_{r\rightarrow\infty}\,.
\end{equation}
The moment of inertia of the star thus is defined to be
\begin{eqnarray}
	I=\frac{J}{\Omega}=\frac{r^4}{6\omega}\frac{{\rm d}\omega}{{\rm d}r} \Big|_{r\rightarrow\infty}\,.
\end{eqnarray}
Combining with Eq.~(\ref{eq: omega}), the moment of inertia of the star can be
calculated from
\begin{eqnarray}
	\frac{{\rm d}I(r)}{{\rm d}r}&=&\frac{2}{3}\frac{4\pi r^4(\rho+p)}{1-2m/r}\left(1-\frac{5}{2}\frac{I}{r^3}+\frac{I^2}{r^6}\right)\,,
\end{eqnarray}
with initial condition $I(r=0)=0$.  This equation, similar to the TOV equation,
can be regarded as the corresponding equation in the Newtonian theory with some
correction terms that come from GR.

\begin{figure}[t]
	\centering
	\includegraphics[width=8cm]{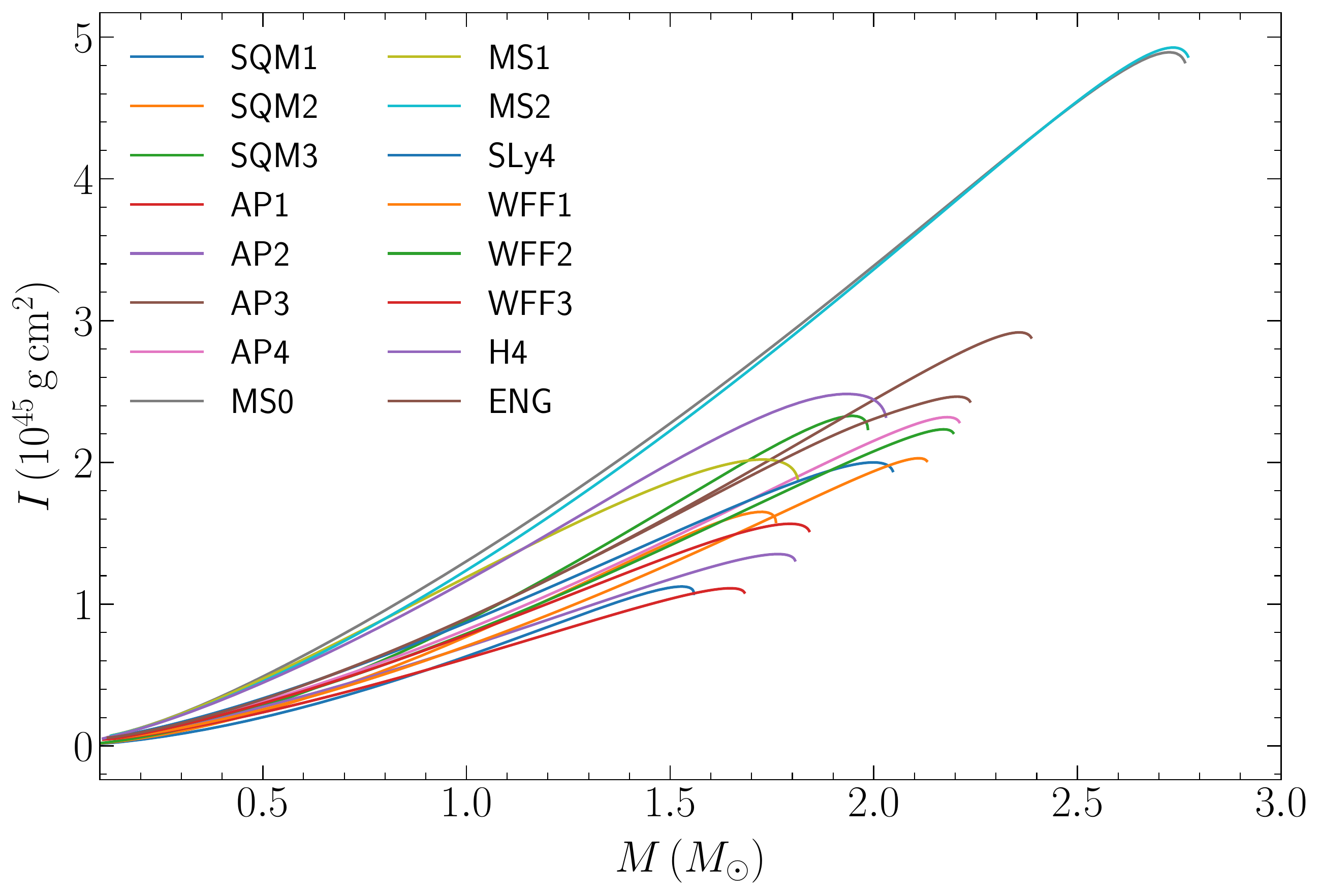}
	\caption{Moment of inertia of NSs with different EOSs (figure courtesy of
	Y. Gao). Unstable solutions are cut off as in Fig.~\ref{fig: M-R}.
	\label{fig: MoI}}
\end{figure}

In Fig.~\ref{fig: MoI} we show the NS moment of inertia with some EOSs.  It is
relatively hard to measure the NS moment of inertia directly, because there is
no leading order effect caused by the current in Newtonian gravity.  Some
indirect measurements come from modeling the X-ray profile of X-ray pulsars and
combining with proper universal relations~\cite{Silva:2020acr}.  Notably, as
explained in Sec.~\ref{sec:doublepulsar}, the observations of the double pulsar
binary starts to constrain the moment of inertia through the spin-orbit
coupling~\cite{Kramer:2021jcw}.  Measuring the moment of inertia independently
is valuable in gravity tests as it relates to the universal relationship that
will be discussed later.

There are some other quantities that describe the NS structure and can be
measured today. One example that will be used later is the tidal Love number of
NSs~\cite{Hinderer:2007mb}.  This quantity describes how easily an object can be
tidally deformed and it can be measured from the GW signal of colliding
NSs~\cite{LIGOScientific:2017vwq}.  At the early stage of the inspiral, the
motion of the binary system is dominated by the point-mass dynamics, which is
similar to, but with higher-order corrections, what we discussed in
Sec.~\ref{sec:obsandtiming}.  Near the end of the inspiral, the non-uniform
gravity field of the companion will cause both stars to be tidally deformed, and
the internal degrees of freedom of the stars begin to influence the GW signal. 
The observation of the famous event GW170817 set an upper limit on the NS tidal
Love number that $\tilde{\rm \Lambda}\lesssim 800$, which excludes some EOSs
that are too stiff~\cite{LIGOScientific:2017vwq}.

There is a problem that comes up when we want to test strong-field gravity with
NS structures: the uncertainties in the EOS~\cite{Shao:2019gjj, Shao:2022koz}.
Nuclear physics in the laboratory cannot tell us which EOS is correct yet.
Thus, for example, the mass-radius relation of NSs depends not only on the
theories of gravity but also on the choice of EOSs.  If a measurement tells us
the mass and radius of a NS, we can equally conclude that in GR, some EOSs are
consistent with this measurement, or for a specific EOS, some gravity theories
can pass the test. Thus, in order to constrain gravity theories, one needs to
consider various reasonable EOSs, or one has to break the degeneracy between
gravity theories and nuclear physics. 

One way to break such a degeneracy is to find relations that are not sensitive
to the EOS but still depend on the gravity theory. In GR, such relations do
exist. One famous example of such relations is the ``I-Love-Q'' relation between
the dimensionless moment of inertia, the tidal Love number and the quadrupole
moment of NSs (see, e.g.\ Fig.~1 in Ref.~\cite{Yagi:2013awa}). The deviations
from the common relation caused by variations in EOS can be smaller than $1\%$.
As the tidal Love number can be measured by GWs from colliding NSs as  mentioned
before, measuring the moment of inertia of NSs thus becomes very important.

Many efforts have been done to understand why there are such universal relations
existing in GR and many other gravity theories~\cite{Yagi:2016bkt}. The
possible physical picture is some approximate symmetries emergent when the stars
become more and more compact and they are responsible for the
universality~\cite{Yagi:2016bkt}. For example, the isodensity self-similarity
leads to a constant eccentricity profile which will minimize the energy of the
system~\cite{Lai:1993ve}. There are also some mathematical approaches showing
that the ``I-Love'' relation is perturbatively insensitive with respect to
changes in the polytropic index of the EOS around the incompressible limit~\cite{Chatziioannou:2014tha}.

There are many other universal relations known today, for example the relation
between the Love number and the $f$-mode frequency, the no-hair-like relations,
the $I$-$C$ relation and so on~\cite{Yagi:2016bkt, Gao:2023mwu}. Testing strong-field gravity
with universal relations seems promising, though there are still some subtle
issues that need to be studied further~\cite{Shao:2022koz}.  In principle, to
test gravity with universal relations, one needs to measure several quantities
independently for the same NS, for example, the Love number and the moment of
inertial. But usually, different quantities are proper to measure in different
systems, which correspond to different NSs with different masses.  One way to go
through this problem is to measure those quantities for many different NSs with
similar masses~\cite{Shao:2022koz}, for example around $1.4\,{\rm M_\odot}$.
Then one can combine these results to a universal relation by doing an
interpolation on the mass. Another possible method is to establish universal
relations that relate quantities in different masses~\cite{Saffer:2021gak}.

\section{Tests of Alternative Gravity Theories}
\label{sec:timing}

Though Einstein's GR has passed all present tests with flying colors, there is
still great enthusiasm for modified gravity theories~\cite{Will:2018bme,
Berti:2015itd, Wex:2014nva, Shao:2016ezh}. Those alternative gravity theories
may arise as low-energy effective theories of some quantum field theories or
motivate from dark matter and dark energy problems. As they provide the explicit
form of how can a theory deviate from GR, studying these modified gravity
theories is also important and intuitive in testing gravity. In the following,
we will describe some basic aspects of modified gravity theories by illustrating
two concrete examples: scalar-tensor gravity and massive gravity. 

\subsection{Scalar-tensor theory}

Scalar-tensor gravity theories are the simplest and mathematically well-posed
extension to GR by including some scalar fields that are non-minimally coupled
in the Lagarangian of gravity. A general class of scalar-tensor gravity theories
with one extra scalar field $\Phi$ have the following action in the physical
frame (or the so-called Jordan frame)~\cite{Damour:1992we}
\begin{eqnarray}
	S&=&\frac{c^4}{16\pi G_*}\int\frac{{\rm d}^4x}{c}\sqrt{-\tilde{g}}[F(\Phi)\tilde{R}-Z(\Phi)\tilde{g}^{\mu\nu}\partial_\mu\Phi\partial_\nu\Phi-U(\Phi)]\nonumber\\
	&&+S_m[\psi_m;\tilde{g}_{\mu\nu}]\,,
\end{eqnarray}
where $G_*$ denotes a bare gravitational coupling constant and
$S_m[\psi_m;\tilde{g}_{\mu\nu}]$ is the action of standard matters; $\tilde{R}$
and $\tilde{g}$ are the Ricci scalar and the metric determinant calculated from the
``physical metric'' $\tilde{g}_{\mu\nu}$, respectively. In this kind of theories
the WEP is satisfied because all the matter variables
$\psi_m$ are coupled to the same metric $\tilde{g}_{\mu\nu}$. It is often
convenient to rewrite this action in a canonical form by redefining $\Phi$
and $\tilde{g}_{\mu\nu}$ via~\cite{Damour:2007uf}
\begin{eqnarray}
	g^*_{\mu\nu}&=&F(\Phi)\tilde{g}_{\mu\nu}\,,\\
	\varphi &=&\pm\int{\rm d}\Phi \left[\frac{3}{4}\frac{F'^2(\Phi)}{F^2(\Phi)}+\frac{1}{2}\frac{Z(\Phi)}{F(\Phi)}\right]^{1/2}\,.
\end{eqnarray}
It leads to an action in the Einstein frame
\begin{eqnarray}\label{eq: Einstein frame action}
	S&=&\frac{c^4}{16\pi G_*}\int\frac{{\rm d}^4 x}{c}\sqrt{-g_*}[R_*-2g_*^{\mu\nu}\partial_\mu\varphi\partial_\nu\varphi-V(\varphi)]\nonumber\\
	&&+S_m[\psi_m;A^2(\varphi)g^*_{\mu\nu}]\,.
\end{eqnarray}
Now $R_*$ and $g_*$ are calculated from the new metric $g^*_{\mu\nu}$, and the
potential term
\begin{eqnarray}
	V(\varphi)&=&F^{-2}(\Phi)U(\Phi)\,.
\end{eqnarray}
In this form the matter fields are coupled to the conformal metric
$A^2(\varphi)g^*_{\mu\nu}$ instead of $g^*_{\mu\nu}$ itself, where the conformal
coupling function
\begin{eqnarray}
	A(\varphi)&=&F^{-1/2}(\Phi)\,.
\end{eqnarray}
As the matter fields couple directly to the metric $\tilde{g}_{\mu\nu}$, this
metric is the one that can be measured by laboratory clocks and rods and it is
called the physical metric. While the canonical representation is often more
convenient for calculation as it decouples the spin-0 excitation of $\varphi$
and the spin-2 excitation of $g^*_{\mu\nu}$.

For different choices of the coupling function $A(\varphi)$ and the potential
$V(\varphi)$, one gets different classes of scalar-tensor theories. Note that
one can always work in some appropriate units that the asymptotic value of
$A(\varphi)=1$, which means in these units, the Einstein metric and the physical
metric asymptotically coincide.

The field equation can be derived from Eq.~(\ref{eq: Einstein frame action}) by
taking variations with respect to $g^*_{\mu\nu}$ and $\varphi$~\cite{Xu:2020vbs}
\begin{eqnarray}
	R^*_{\mu\nu}&=&2\partial_\mu\varphi\partial_\nu\varphi+8\pi G_*\left(T^*_{\mu\nu}-\frac{1}{2}T^*g*_{\mu\nu}\right)+\frac{1}{2}g^*_{\mu\nu}V(\varphi)\,,\\
	\Box\varphi &=& -4\pi G_*\alpha(\varphi)T_*+\frac{1}{4}\frac{{\rm d}V(\varphi)}{{\rm d}\varphi}\,,
\end{eqnarray}
where $T^{\mu\nu}_*=2(-g_*)^{-1/2}{\rm \delta}S_m/{\rm \delta}g^*_{\mu\nu}$ and
$\alpha(\varphi)={\rm \partial}\ln A(\varphi)/{\rm \partial}\varphi$. The
quantity $\alpha(\varphi)$ plays the role of the coupling strength between the
scalar field and ordinary matter, and in general, it can be field-dependent.

Since any function $a(\varphi)\equiv \ln A(\varphi)$ can be expanded when the
scalar field is weak as
\begin{eqnarray}\label{eq: coupling expansion}
	a(\varphi)&=& \alpha_0(\varphi-\varphi_0)+\frac{1}{2}\beta_0(\varphi-\varphi_0)^2+\cdots\,,
\end{eqnarray}
it has been shown that, for $V(\varphi)=0$, in the appropriate units that
$a(\varphi_0)=0$, all measurable quantities depend only on the first two
coefficients $\alpha_0$ and $\beta_0$ of the expansion at the 1\,PN level for
weakly self-gravitating objects~\cite{Damour:1992we}. In the PPN formalism~\cite{Will:2018bme}, only
two parameters $\gamma^{\rm PPN}$ and $\beta^{\rm PPN}$ deviate from GR (in
which $\gamma^{\rm PPN}=\beta^{\rm PPN}=1$), and they are given by
\begin{eqnarray}
	\gamma^{\rm PPN}=1-\frac{2\alpha_0^2}{1+\alpha_0^2}\,, \quad \quad
	\beta^{\rm PPN}=1+\frac{\alpha_0^2\beta_0}{2(1+\alpha_0^2)^2}\,.
\end{eqnarray}
One can also find that the effective gravitational constant between two bodies
is $G=G_*(1+\alpha_0^2)$ in the weak field.

A famous theory, which also may be regarded as the simplest case for
scalar-tensor gravity, is the Brans-Dicke theory~\cite{Brans:1961sx}, where
there is a varying gravitational constant.  In the scalar-tensor formalism, the
Brans-Dicke theory is equivalent to
\begin{eqnarray}
	a(\varphi)&=&\alpha_0\varphi+{\rm const.}\,,  \quad \quad V(\varphi)=0\,,
\end{eqnarray}
and in this theory there is a field-independent coupling strength,
$\alpha(\varphi)=\alpha_0$.  Solar system experiments including light deflection, time delay and frequency shift measurements made with the Cassini spacecraft have set limit on the coupling strength of the scalar field that
$\alpha_0^2\lesssim 1\times10^{-5}$~\cite{Reasenberg:1979ey,Robertson1991,Bertotti:2003rm}, which tightly constrains the deviations from GR in the
weak-field region.

Things can be much different for systems including strongly self-gravitating
objects like NSs. It was found that a wide class of scalar-tensor theories can
exhibit nonperturbative behaviors called spontaneous scalarization for strong-field objects~\cite{Damour:1993hw, Doneva:2022ewd}.  As those theories can pass
all the weak-field gravity tests while providing large deviations from GR in the
strong-field region and still keep good mathematical behaviors, they largely
enhance the interests in studying such theories.

For strongly self-gravitating objects, one can define the effective scalar
charge of an object $A$ by
$	\alpha_A = {{\rm \partial }\ln m_A}/{{\partial}\varphi_0}$,
or equivalently by the asymptotic behavior of the scalar field~\cite{Damour:1992we}
	$\varphi=\varphi_0-{G_* m_A \alpha_A}/{r}+O\left({1}/{r^2}\right)$.
The Newtonian gravitational coupling between two bodies $A$ and $B$ can be
described by the effective coupling constant,
	$G_{AB}=G_*(1+\alpha_A\alpha_B)$.

A simple theory that can have spontaneous scalarization is called the
Damour-Esposito-Farèse (DEF) theory, which is widely
studied~\cite{Damour:1993hw, Damour:2007uf}.  It can be described by
	$A(\varphi)=e^{\beta \varphi^2/2}$ and	$V(\varphi)=0$,
and the asymptotic value $\varphi_0$ of $\varphi$ here we set to be zero. With
arbitrary choice of the asymptotic value $\varphi_0$, the DEF theory is just a
simple choice with $\alpha_0=\beta\varphi_0$, $\beta_0=\beta$ in Eq.~(\ref{eq:
coupling expansion}). In this theory, the coupling strength
$\alpha(\varphi)=\beta\varphi$ is a linear function of $\varphi$.

The name spontaneous scalarization used by Damour and Esposito-Far\`ese comes from
the analogy to the phenomenon of spontaneous magnetization of a
ferromagnet~\cite{Damour:1996ke, Sennett:2017lcx}. In the DEF theory, the
trivial case that $\varphi=0$ is always a solution to the field equations. In
fact this solution is identical with that in GR.  In the weak-field region, as
we have set $\varphi_0=0$, the DEF theory reduces to GR so it can pass any
weak-field gravity tests that GR can pass.  But for a strongly self-gravitating
body, a large scalar charge may be produced spontaneously, like a ferromagnet
will have spontaneous magnetization at low temperature.  The solution with a
non-vanish scalar charge is energetically favored when the star's baryon
mass exceeds the critical mass~\cite{Damour:1993hw, Doneva:2022ewd}.

One can also understand this phenomenon by analyzing the instability of the
perturbation in the scalar field~\cite{Ramazanoglu:2016kul}. To the linear order
in $\varphi$, the field equations become
\begin{eqnarray}
	R^*_{\mu\nu}&=&8\pi\left(\tilde{T}_{\mu\nu}-\frac{1}{2}g^*_{\mu\nu}\tilde{T}\right)\,,\\
	\Box \varphi &=& -4\pi \tilde{T}\beta \varphi\,.\label{eq: linear phi}
\end{eqnarray} 
For matter described by a perfect fluid, we have
$\tilde{T}=\tilde{\rho}+3\tilde{p}$, which is negative for ordinary matter. Thus
the right-hand side of Eq.~(\ref{eq: linear phi}) is negative for a negative
$\beta$, and the scalar field can suffer the tachyonic instability for a
sufficiently negative $\beta$. Interestingly, it has been shown in some works
that for highly compact NSs with certain EOSs, $\tilde{T}$ can be positive,
which leads to a spontaneous scalarization for $\beta>0$~\cite{Mendes:2016fby}.

For those scalar-tensor theories with spontaneous scalarization, the weak-field
tests cannot constrain their parameter space efficiently, but they still can be
constrained from strong-field tests. In scalar-tensor theories, the energy loss
of a system can be carried by both the spin-2 GWs and spin-0 scalar waves to
infinity.  For those theories with spontaneous scalarization, NSs can carry a
large number of scalar charges, which will cause a much enhanced gravitational
dipolar radiation for a scalarized NS in a binary.  To the leading order, it
will contribute an additional decay rate of the binary orbital period
via~\cite{Damour:1996ke}
\begin{eqnarray}
	\dot{P}_b^{\rm dipole}=-\frac{2\pi G_*}{c^3}(1+e^2/2)(1-e^2)^{-5/2}\frac{2\pi}{P_b}\frac{m_A m_B}{M}(\alpha_A-\alpha_B)^2\,.
\end{eqnarray}
For some binary pulsars, while a NS has a significant scalar charge, its companion, for example a WD, can
only take a vanishingly small scalar charge $\alpha_B\rightarrow 0$ as it is a
weak-field object. Therefore, NS-WD binaries can be the most sensitive probe to
constrain the parameter space of spontaneous scalarization~\cite{Freire:2012mg}.
There is also study showing that for NS-NS binaries with significant difference in
the masses of the components can also be excellent
laboratories~\cite{Zhao:2022vig}.  Besides that, it has been proven that the
no-hair theorem still holds in many scalar-tensor theories, which means that the
BHs in those theories cannot carry a scalar charge. So that although the binary
BHs in those theories will not excite dipole radiation, the NS-BH binaries can
still provide abundant information about the underlying gravity
theory~\cite{Liu:2014uka, Shao:2014wja}.

\begin{figure}[t]
	\centering
	\includegraphics[width=8cm]{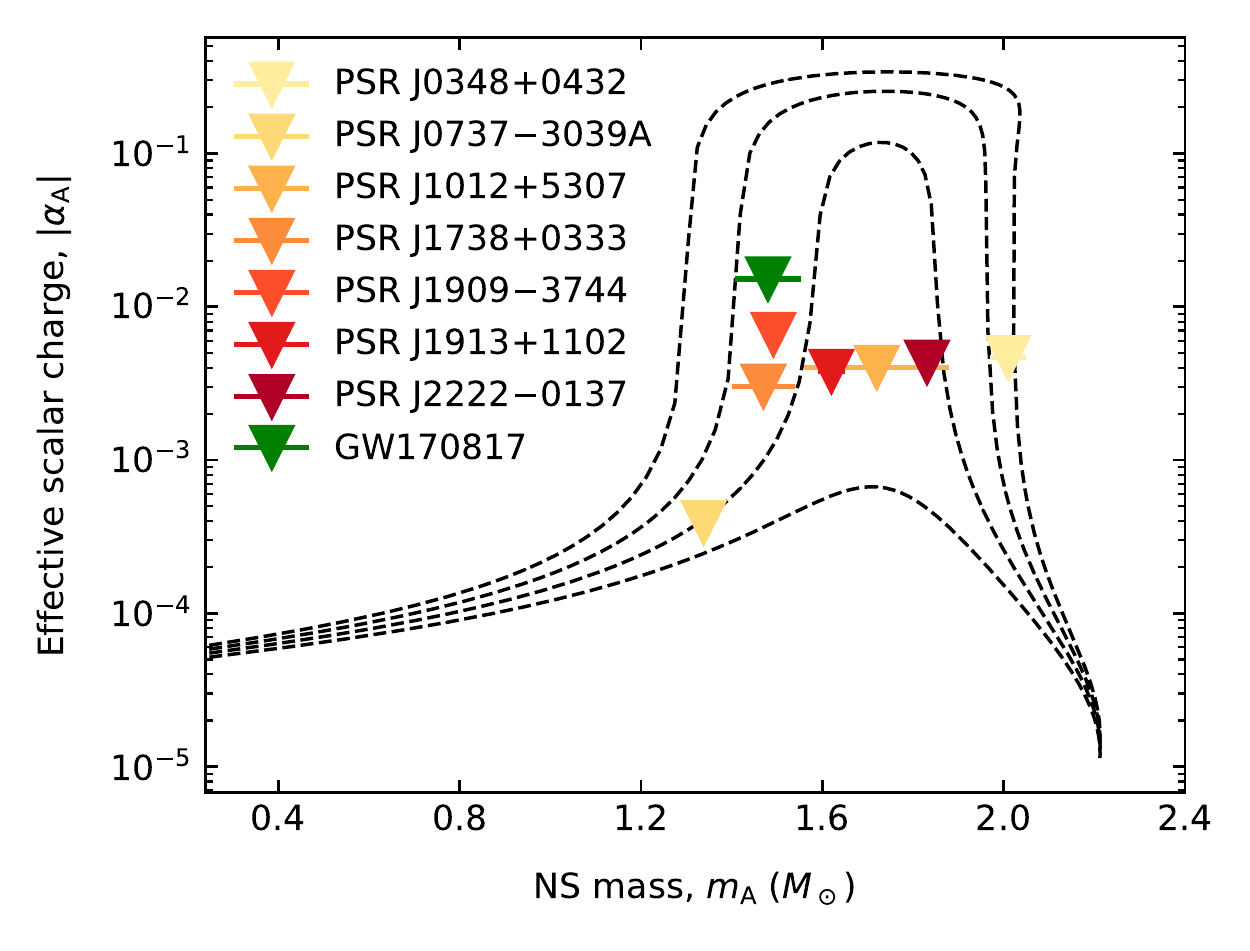}
	\caption{\label{fig: limit on scalar charge} The dashed black curves show
	the effective scalar charge in the DEF theory with $|\alpha_0|=10^{-5}$ and
	$\beta_0=-4.8,\,-4.6,\,-4.4,\,-4.2$ from top to bottom repectively. The EOS
	is assumed to be AP4. The observational bounds from different binary pulsars
	 and a GW event are denoted by the triangles~\cite{Shao:2022izp}.}
\end{figure}

Figure~\ref{fig: limit on scalar charge} illustrates the constraints on the
effective scalar charges from the gravitational dipole radiation of seven binary
pulsar systems and one GW event~\cite{Shao:2022izp, Zhao:2022vig}. Note that in
this figure, a specific EOS, AP4, is used to draw the curves. In principle, the
uncertainty in the NS EOS will entangle with the gravity test. Nevertheless,
with enough well-measured binary pulsar systems, one can set a limit on the
whole mass range for NSs, and studies have shown that for each reasonable EOS,
the parameter space for spontaneous scalarization in the DEF theory is now very
small. In Refs.~\cite{Shao:2017gwu, Zhao:2019suc, Guo:2021leu, Zhao:2022vig}, by using the Bayesian
parameter estimation,  the possibility of the spontaneous scalarization for an
effective coupling larger than $10^{-2}$ for DEF theory is basically closed, no
matter what the underlying EOS of NS is.

As the parameter space for spontaneous scalarization in the initial DEF theory
is largely constrained by pulsar timing observation, a natural way to avoid
the constraints while keeping the phenomenon of spontaneous scalarization is to
consider massive scalar-tensor theories~\cite{Ramazanoglu:2016kul,
Doneva:2022ewd}. For example one can consider the DEF theory but with
$V(\varphi)=2m^2\varphi^2$, where $m$ is the mass of the scalar field. The
linearized field equation of the scalar field now becomes
\begin{eqnarray}
	\Box \varphi &=& (m^2-4\pi \tilde{T}\beta )\varphi\,.\label{eq: linear phi}
\end{eqnarray}
It shows that the mass term of the scalar field will suppress the existence of
spontaneous scalarization~\cite{Ramazanoglu:2016kul}.  Nevertheless, the
spontaneous scalarization can still happen for massive scalar-tensor theories.
But with a mass term, the asymptotic value of the scalar field is suppressed
exponentially in a Yukawa fashion, which means by definition the scalar charge
of a scalarized NS vanishes.  Thus no significant dipole radiation will go to
infinity. What is more, when the Compton wavelength of the scalar field is
smaller than the orbital separation of the binary, the modification with respect
to GR in the orbital dynamics will also be suppressed exponentially, which
causes the pulsar timing observation hardly setting strong constraints in some
parameter space of massive scalar-tensor theories. Fortunately, this kind of
theory can still be constrained via the tidal deformability measurement from GWs
(see e.g.\ Refs.~\cite{Xu:2020vbs, Hu:2021tyw}). This is true for the reason
that a scalarized NS can have large difference in radius compared to that in GR,
while the tidal deformability is very sensitive to the star's radius. In this
sense, combining pulsar timing observation and GW observation can probe
a large parameter space of scalar-tensor theories~\cite{Shao:2017gwu}.

In recent years, some variants of scalar-tensor theories triggered great
enthusiasm as they can provide solutions for scalarized BHs, in contrast to the
no-hair theorem (see Ref.~\cite{Doneva:2022ewd} for a review). One example of
those theories is the scalar-Gauss-Bonnet theory, which includes a topological
Gauss-Bonnet term,
\begin{eqnarray}
	\mathcal{G}&=&R_{\alpha\beta\gamma\delta}R^{\alpha\beta\gamma\delta}-4R_{\alpha\beta}R^{\alpha\beta}+R^2\,,
\end{eqnarray}
which is coupled to the scalar field~\cite{Doneva:2017bvd,
Silva:2017uqg, Xu:2021kfh}.
Scalarization of NSs in such kinds of theories can be very different from those
in the DEF theories. It also provides a new field where observations of compact
objects, including NSs and BHs, are crucial to revealing the
strong-field information of gravity. More examples of other variants of
scalar-tensor theories that pass current binary-pulsar tests can be found in
Ref.~\cite{Doneva:2022ewd}.

As already discussed, the theories of gravity can sometimes degenerate with the
EOS of NSs. One way to break the degeneracy is to find those universal relations
that do not depend on the EOS. We have shown that such universal relations do
exist in GR, and in fact, this can still be true for some modified gravity. For
some theories without spontaneous scalarization, universal relations in GR, like
the I-Love-Q relation, are still valid. That means in  modified
theories, these relations also only weakly depend on the EOS of
NSs~\cite{Yagi:2016bkt}, but they can be very different from their GR
counterpart. Things can be more interesting when spontaneous scalarization kicks
in. The universal relations in GR can be EOS dependent for those theories with
spontaneous scalarization, because in general, the starting point of the
spontaneous scalarization is different for different EOSs. Once a NS is
spontaneously scalarized, the relations can differ from their GR counterparts
largely. Thus for theories with spontaneous scalarization, we can see that the
relations for different EOSs branching from the GR one from different starting
points. For some theories, like the DEF theory, those different relations for
different EOSs will tend to form a new universal relation different from GR when
the scalarization becomes stronger. However for some theories like the massive
Gauss-Bonnet theory, this may or may not happen depending on the theoretical
parameters~\cite{Xu:2021kfh}.

\subsection{Massive gravity theory}

From a modern particle physics perspective, the related quantum particle for
gravity in GR is a massless spin-2 particle, the so-called  graviton.
However, massive gravity theories, which maintain the notion that gravity is
propagated by a spin-2 particle but consider this particle to be massive, have
aroused wide interest. Such a massive gravity theory may be used to explain the
acceleration of the Universe and dark matter~\cite{Hinterbichler:2011tt}.
Probing the upper limit of the graviton mass is a basic topic in physics.
Constraints from the pulsar timing technique may come from two aspects: one is
the radiative tests from binary pulsars, and the other is the orbital motion of
a pulsar-supermassive BH system.

Unlike the dispersion relation tests from the LIGO/Virgo/KAGRA GW
events~\cite{LIGOScientific:2021sio}, the constraints on the graviton mass are
generally theory dependent in pulsar timing, therefore we shall introduce some
specific massive gravity theories. One of the examples is the cubic Galileon
theory, which is a simple model with the Vainshtein screening
mechanism~\cite{deRham:2012fw}. The action including matter is 
\begin{eqnarray}
	S&=&\int {\rm d}^4 x\left[-\frac{1}{4}h^{\mu\nu}(\mathcal{E}h)_{\mu\nu}+\frac{h^{\mu\nu}T_{\mu\nu}}{2M_{\rm Pl}}-\frac{3}{4}(\partial \pi_s)^2\left(1+\frac{1}{3\Lambda^3}\Box\pi_s\right)+\frac{\pi_s T}{2 M_{\rm Pl}}\right]\,,\nonumber\\
	&&
\end{eqnarray}
where $h_{\mu\nu}=g_{\mu\nu}-\eta_{\mu\nu}$,
$(\mathcal{E}h)_{\mu\nu}=-\frac{1}{2}\Box h_{\mu\nu}+\cdots$ and $T$ is the
trace of the stress-energy tensor. In the action, $M_{\rm Pl}$ is the Planck
mass, and the strong coupling scale of the Galileon sector $\Lambda$ is related
to the graviton mass $m_g$ via $\Lambda^3 = m_g^2 M_{\rm Pl}$. The field equation for
$\pi_s$ and $h_{\mu\nu}$ are~\cite{deRham:2012fw}
\begin{eqnarray}
	\frac{1}{M_{\rm Pl}}T_{\mu\nu}&=&-\frac{1}{2}\Box h_{\mu\nu}\,,\\
	\frac{1}{2 M_{\rm Pl}}T&=&\partial_{\mu}\left[-\frac{3}{2}\partial^\mu\pi_s\left(1+\frac{1}{3\Lambda^3}\Box \pi_s\right)+\frac{1}{4\Lambda^3}\partial^{\mu}(\partial\pi_s)^2\right]\,.
\end{eqnarray}
For a static point source with mass $M$, the solution for $\pi$ can be written
as $\nabla \pi_s(r)=\hat{r}E(r)$, where
\begin{eqnarray}
	E_{\pm}(r)=\frac{\Lambda^3}{4r}\left[\pm\sqrt{9r^4+\frac{32}{\pi}r_\star^3 r}-3r^2\right]\,,
\end{eqnarray}
and the Vainshtein radius associated with the mass $M$ is
	$r_\star = \left({M}/{16 \Lambda^3 M_{\rm pl}}\right)^{1/3}= \left({M}/{16 m_g^2 M_{\rm pl}^2}\right)^{1/3}$.
Within the radius $r_\star$, the fifth force from the scalar field is strongly
suppressed, and the theory reduces to the canonical gravity. With such a
screening mechanism, this theory can avoid the stringent constraints from the
Solar system while it still can deviate from GR at a large scale, thus providing
changes to the cosmological evolution.

Though the screening mechanism tends to suppress the deviation from GR at
high-density regions, the cubic Galileon theory still predicts difference in
gravitational radiation~\cite{deRham:2012fw}. For a binary system with a
characteristic velocity $v$, the gravitational radiation is less suppressed by a
factor of $\sim v^{3/2}$ compared with its fifth-force counterpart. As shown by
de Rham et al.~\cite{deRham:2012fw}, the extra radiative channels in these
theories include monopolar radiation, dipolar radiation and quadrupolar
radiation, and for different systems, for example binary pulsar systems with
different orbital periods and orbital eccentricities, the dominate radiation
channel can be different~\cite{Shao:2020fka}. Recent study with 14 well-timed
binary pulsar systems has set a tight constraint on the graviton mass in the
cubic Galileon theory, $m_g< 2\times 10^{-28}\,{\rm eV/c^2}$, at 95\% confidence
level~\cite{Shao:2020fka}.

Another example we discuss here is the so-called Yukawa gravity, which is a
widely used phenomenological form that contains the possible deviations from the
Newtonian gravity via a modified potential,
\begin{eqnarray}\label{eq: Yukawa Potential}
	\Phi(r)&=&-\frac{GM}{(1+\alpha)r}\left(1+\alpha e^{-r/{\rm\Lambda}_{\rm Y}}\right)\,.
\end{eqnarray}
Part of the Newtonian potential is suppressed exponentially, with a strength of
$\alpha$ and a length scale of ${\rm\Lambda}_{\rm Y}$. For $\alpha=0$ or ${\rm\Lambda}_{\rm Y}=\infty $,
the Yukawa potential will reduce to the Newtonian one, and the graviton mass
here can be expressed as $m_g=h/{\rm\Lambda}_{\rm Y}$.  Such a Yukawa gravity can naturally
arise in the Newtonian limit of $f(R)$ gravity with a general
action~\cite{CAPOZZIELLO}
\begin{eqnarray}
	S&=&\int {\rm d}^4 x\sqrt{-g}\left[f(R)+\frac{16\pi G}{c^4}\mathcal{L}_m\right]\,.
\end{eqnarray}
Assuming that the function $f(R)$ is Taylor expandable and take the $O(2)$
approximation, $f(R)=f_0+f_1 R +f_2 R^2 +\cdots$ with constants $f_i$ $(i=1,2,3)$, the gravitational potential
defined via $g_{tt}=1+2\Phi$ will take the form of Eq.~(\ref{eq: Yukawa
Potential}), and the Newtonian potential is obtained only when $f(R)=f_1 R$.

Tests of the Yukawa gravity have been carried out at a wide range of length
scales with different systems and methods, including terrestrial laboratory
experiments~\cite{Niebauer:1987ua}, LAser GEOdynamic Satellite
(LAGEOS)~\cite{Peron:2014pba}, Lunar Laser Ranging (LLR)~\cite{Williams:2005rv},
Solar system planetary orbits~\cite{Talmadge:1988qz}, S2 star
orbit~\cite{Zakharov:2016lzv} and observations of elliptical
galaxies~\cite{Napolitano:2012fp}. The reason for taking experiments at such a
wide coverage is that for the Yukawa gravity theory, only when the system scale
$L$ is similar to the length scale $\Lambda_{\rm Y}$ it shows the largest deviation from
Newtonian gravity. Thus timing a pulsar orbiting around the supermassive BH in
our galaxy center may provide a unique opportunity to test the Yukawa gravity
theory at a new length scale~\cite{Dong:2022zvh}. For a pulsar orbit with a
semi-major axis $a$ and an orbital eccentricity $e$, it will experience an
additional periastron advance from the Yukawa potential at the rate
of~\cite{Zakharov:2018cbj}
\begin{eqnarray}\label{eq: Yukawa advance}
	\dot{\omega}&=&\frac{1}{P_b}\frac{\pi \alpha \sqrt{1-e^2}}{1+\alpha}\frac{a^2}{{\rm\Lambda}_{\rm Y}^2}\,,
\end{eqnarray}
to the first order in $a/{\rm\Lambda}_{\rm Y}$. If the pulsar orbit has a semi-major axis
$a\sim {\rm\Lambda}_{\rm Y}$, periodic effects such as the deformation of the orbit may also
be detectable by pulsar timing.

\begin{figure}[t]
	\centering
	\includegraphics[width=8cm]{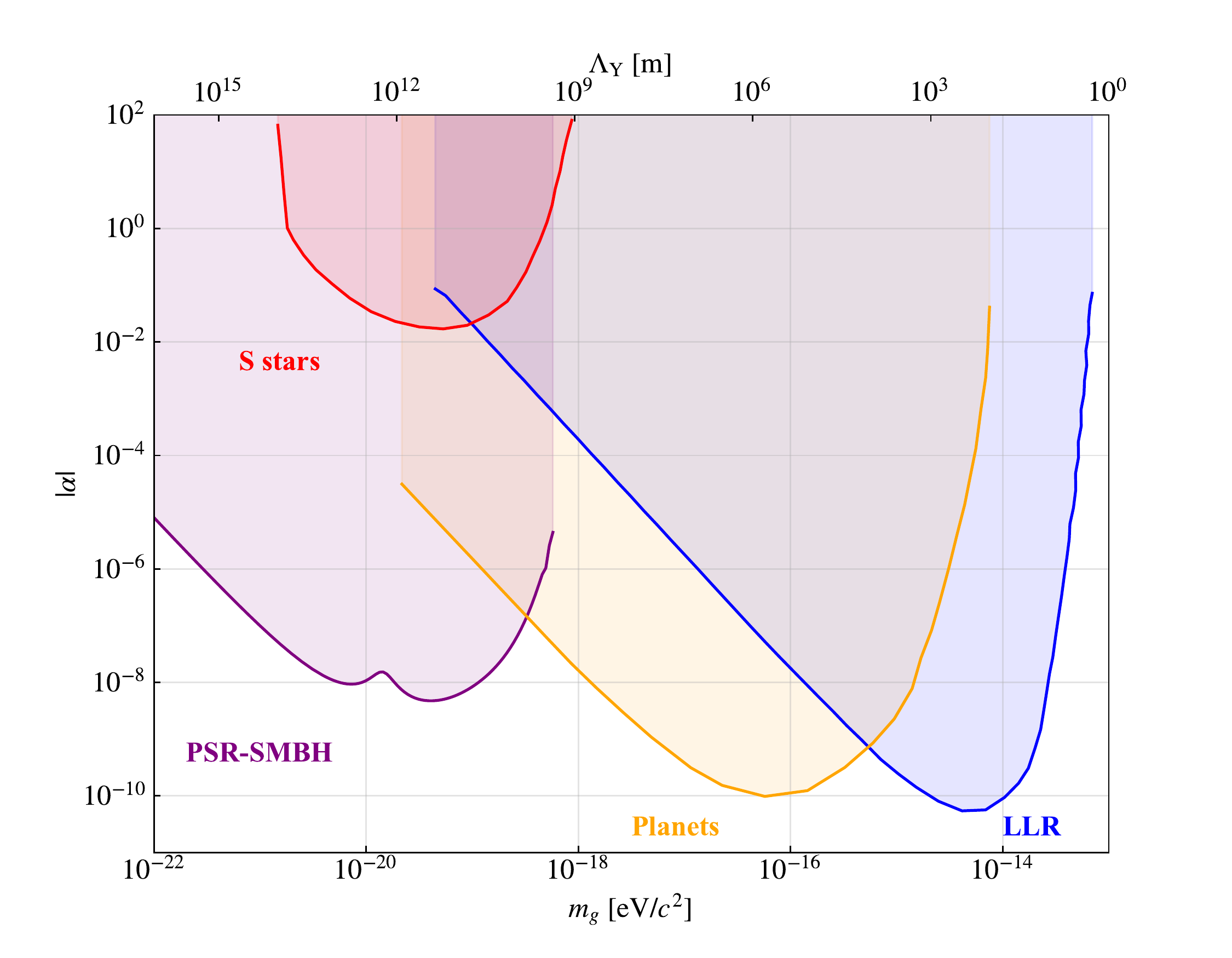}
	\caption{\label{fig: Yukawa constrains} Constraints on the strength of the
	Yukawa interaction $\alpha$ as a function of the graviton mass $m_g$ from different experiments (figure courtesy of Y.
	Dong)~\cite{Dong:2022zvh}. Shaded regions show the excluded parameter space
	at the 95\% confidence level.}
\end{figure}

Figure~\ref{fig: Yukawa constrains} shows the projected constraints from a S2-like like pulsar, on the
strength $\alpha$ at 95\% confidence level as a function of
$m_g$~\cite{Dong:2022zvh}, from where one can see that the
pulsar-supermassive BH system can contribute a unique test of this
theory that is not covered by other experiments. For a given strength of the
Yukawa interaction $\alpha$ one also can obtain the related constraint on the
graviton mass, and from Eq.~(\ref{eq: Yukawa advance}) one can see that for
$\alpha$ changing from 1 to $\infty$, the periastron advance rate only changes by a
factor of 2, thus the constraints on $m_g$ will not change too much.
Simulation results by Dong et al.~\cite{Dong:2022zvh} showed that the expected
constraints on graviton mass can reach $m_g\lesssim 10^{-24}\,{\rm eV/c^2}$ for
a 20-yr observation of a S2-like pulsar around Sagittarius A$^\star$.

\section{Summary}
\label{sec:summary}
In this chapter, we present a pedagogical introduction to some basic aspects of pulsar timing and using it in testing gravity theories. As pulsar timing provides very high accurate measurements of a strongly self-gravitating object in orbit, it can put strict constraints in the strong-field region of gravity. Moreover, the observational precision of the pulsar-timing parameters generally improves with the observational time span $T_{\rm obs}$. For example, the precision in the orbital decay parameter, $\dot{P}_b$, improves as $T_{\rm obs}^{-5/2}$. Thus long-term observations of interesting systems like the Hulse-Taylor pulsar and the double pulsar can provide us with new results from time to time until the precision is dominated by other effects (e.g.\ the Galactic potential model). Furthermore, the sensitivity of radio telescope is improving at the same time, so the real improvement of timing parameters is faster than the theoretical power law predictions. New telescopes like the FAST telescope in China~\cite{Lu:2019gsr} and the next-generation radio telescope, the Square Kilometre Array in South Africa and Australia~\cite{Weltman:2018zrl} can provide plenty of new results. New systems like pulsar-BH binaries or pulsars near the Galactic center may be discovered, and they may lead to new tests of gravity and give better results~\cite{Liu:2011ae, Liu:2014uka, Shao:2014wja}.

As we mentioned in the chapter, combining pulsar timing observation with other observations can probe a large theory space of both gravity theories and nuclear physics. GW observations~\cite{LIGOScientific:2021sio} and BH shadows~\cite{EventHorizonTelescope:2022xqj, EventHorizonTelescope:2020qrl, Shao:2022wdn} all provide unique information of gravity in different gravity regimes. For GWs, with the third generation of detectors that are under construction, almost all binary NS mergers in the Universe can be detected, and the BH shadows and X-ray tests provide the possibility of measuring the BH metric in the near zone. Combining with proper universal relations, one can break the degeneracy between gravity theories and nuclear physics, thus giving EOS-unbiased tests of gravity.



\begin{acknowledgement}
We thank Yiming Dong and Yong Gao for providing figures.
This work was supported by the National SKA Program of China (2020SKA0120300),
the National Natural Science Foundation of China (11975027, 11991053, 12203072, 11721303),
and the Max Planck Partner Group Program funded by the Max Planck Society.  
X.M.\ is supported by the Cultivation Project for FAST Scientific Payoff and Research Achievement of CAMS-CAS.
\end{acknowledgement}


\normalem
\bibliographystyle{abbrv}
\bibliography{refs}

\begin{thebibliography}{100}

\bibitem{LIGOScientific:2017vwq}
B.~P. Abbott et~al.
\newblock {GW170817: Observation of Gravitational Waves from a Binary Neutron
  Star Inspiral}.
\newblock {\em Phys. Rev. Lett.}, 119(16):161101, 2017.

\bibitem{LIGOScientific:2021sio}
R.~Abbott et~al.
\newblock {Tests of General Relativity with GWTC-3}.
\newblock {\em arXiv e-prints}, page arXiv:2112.06861, 2021.

\bibitem{EventHorizonTelescope:2022xqj}
K.~Akiyama et~al.
\newblock {First Sagittarius A* Event Horizon Telescope Results. VI. Testing
  the Black Hole Metric}.
\newblock {\em Astrophys. J. Lett.}, 930(2):L17, 2022.

\bibitem{Antoniadis:2013pzd}
J.~Antoniadis et~al.
\newblock {A Massive Pulsar in a Compact Relativistic Binary}.
\newblock {\em Science}, 340:6131, 2013.

\bibitem{Archibald:2018oxs}
A.~M. Archibald, N.~V. Gusinskaia, J.~W.~T. Hessels, A.~T. Deller, D.~L.
  Kaplan, D.~R. Lorimer, R.~S. Lynch, S.~M. Ransom, and I.~H. Stairs.
\newblock {Universality of Free Fall from the Orbital Motion of a Pulsar in a
  Stellar Triple System}.
\newblock {\em Nature}, 559(7712):73--76, 2018.

\bibitem{1970Natur_backer}
D.~C. Backer.
\newblock {Pulsar Nulling Phenomena}.
\newblock {\em Nature}, 228(5266):42--43, 1970.

\bibitem{Bagchi_2018}
M.~Bagchi.
\newblock {Prospects of Constraining the Dense Matter Equation of State from
  Timing Analysis of Pulsars in Double Neutron Star Binaries: The Cases of PSR
  J0737-3039A and PSR J1757-1854}.
\newblock {\em Universe}, 4(2):36, 2018.

\bibitem{1982ApJ_Bartel}
N.~Bartel, D.~Morris, W.~Sieber, and T.~H. Hankins.
\newblock {The Mode-Switching Phenomenon in Pulsars}.
\newblock {\em Astrophys. J.}, 258:776--789, 1982.

\bibitem{Bauswein:2017aur}
A.~Bauswein and N.~Stergioulas.
\newblock {Semi-Analytic Derivation of the Threshold Mass for Prompt Collapse
  in Binary Neutron Star Mergers}.
\newblock {\em MNRAS}, 471(4):4956--4965, 2017.

\bibitem{Berti:2015itd}
E.~Berti et~al.
\newblock {Testing General Relativity with Present and Future Astrophysical
  Observations}.
\newblock {\em Class. Quant. Grav.}, 32:243001, 2015.

\bibitem{Bertotti:2003rm}
B.~Bertotti, L.~Iess, and P.~Tortora.
\newblock {A Test of General Relativity Using Radio Links with the Cassini
  Spacecraft}.
\newblock {\em Nature}, 425:374--376, 2003.

\bibitem{Blandford:1976}
R.~Blandford and S.~A. Teukolsky.
\newblock {Arrival-Time Analysis for a Pulsar in a Binary System}.
\newblock {\em Astrophys. J.}, 205:580--591, 1976.

\bibitem{Bondi:1964zz}
H.~Bondi.
\newblock {Massive Spheres in General Relativity}.
\newblock {\em Proc. Roy. Soc. Lond. A}, 282:303--317, 1964.

\bibitem{Brans:1961sx}
C.~Brans and R.~H. Dicke.
\newblock {Mach's Principle and a Relativistic Theory of Gravitation}.
\newblock {\em Phys. Rev.}, 124:925--935, 1961.

\bibitem{Breton2008}
R.~P. Breton, V.~M. Kaspi, M.~Kramer, M.~A. McLaughlin, M.~Lyutikov, S.~M.
  Ransom, I.~H. Stairs, R.~D. Ferdman, F.~Camilo, and A.~Possenti.
\newblock {Relativistic Spin Precession in the Double Pulsar}.
\newblock {\em Science}, 321:104--107, 2008.

\bibitem{Burgay:2003Natur}
M.~Burgay et~al.
\newblock {An Increased Estimate of the Merger Rate of Double Neutron Stars
  From Observations of a Highly Relativistic System}.
\newblock {\em Nature}, 426:531--533, 2003.

\bibitem{Caleb:2022}
M.~Caleb et~al.
\newblock {Discovery of a radio-emitting neutron star with an ultra-long spin
  period of 76 s}.
\newblock {\em Nature Astronomy}, 6:828--836, 2022.

\bibitem{CAPOZZIELLO}
S.~Capozziello, A.~Stabile, and A.~Troisi.
\newblock {A General Solution in the Newtonian Limit of f(R)-Gravity}.
\newblock {\em Mod. Phys. Lett. A}, 24:659--665, 2009.

\bibitem{Chatziioannou:2014tha}
K.~Chatziioannou, K.~Yagi, and N.~Yunes.
\newblock {Toward Realistic and Practical No-Hair Relations for Neutron Stars
  in the Nonrelativistic Limit}.
\newblock {\em Phys. Rev. D}, 90(6):064030, 2014.

\bibitem{Clifton_2012}
T.~Clifton, P.~G. Ferreira, A.~Padilla, and C.~Skordis.
\newblock {Modified Gravity and Cosmology}.
\newblock {\em Phys. Rept.}, 513:1--189, 2012.

\bibitem{Cordes:2002astro}
J.~M. Cordes and T.~J.~W. Lazio.
\newblock {NE2001. 1. A New Model for the Galactic Distribution of Free
  Electrons and its Fluctuations}.
\newblock {\em arXiv e-prints}, pages astro--ph/0207156, 2002.

\bibitem{1990AJ_Cordes}
J.~M. Cordes, J.~M. Weisberg, and T.~H. Hankins.
\newblock {Quasiperiodic Microstructure in Radio Pulsar Emission}.
\newblock {\em Astron. J.}, 100:1882, 1990.

\bibitem{Damour:2007uf}
T.~Damour.
\newblock {Binary Systems as Test-Beds of Gravity Theories}.
\newblock In {\em {6th SIGRAV Graduate School in Contemporary Relativity and
  Gravitational Physics: A Century from Einstein Relativity: Probing Gravity
  Theories in Binary Systems}}, 2007.

\bibitem{Damour1985}
T.~Damour and N.~Deruelle.
\newblock {General Relativistic Celestial Mechanics of Binary Systems. I. The
  Post-Newtonian Motion.}
\newblock {\em Ann. Inst. H. Poincar{\'e} Phys. Th{\'e}or.}, 43(1):107--132,
  1985.

\bibitem{Damour1986}
T.~Damour and N.~Deruelle.
\newblock {General Relativistic Celestial Mechanics of Binary Systems. II. The
  Post-Newtonian Timing Formula}.
\newblock {\em Ann. Inst. H. Poincar{\'e} Phys. Th{\'e}or.}, 44(3):263--292,
  1986.

\bibitem{Damour:1992we}
T.~Damour and G.~Esposito-Far{\`e}se.
\newblock {Tensor Multiscalar Theories of Gravitation}.
\newblock {\em Class. Quant. Grav.}, 9:2093--2176, 1992.

\bibitem{Damour:1993hw}
T.~Damour and G.~Esposito-Far{\`e}se.
\newblock {Nonperturbative Strong Field Effects in Tensor-scalar Theories of
  Gravitation}.
\newblock {\em Phys. Rev. Lett.}, 70:2220--2223, 1993.

\bibitem{Damour:1996ke}
T.~Damour and G.~Esposito-Far{\`e}se.
\newblock {Tensor-scalar Gravity and Binary Pulsar Experiments}.
\newblock {\em Phys. Rev. D}, 54:1474--1491, 1996.

\bibitem{Damour:2009vw}
T.~Damour and A.~Nagar.
\newblock {Relativistic Tidal Properties of Neutron Stars}.
\newblock {\em Phys. Rev. D}, 80:084035, 2009.

\bibitem{Damour:1988}
T.~Damour and G.~Schaefer.
\newblock {Higher Order Relativistic Periastron Advances and Binary Pulsars}.
\newblock {\em Nuovo Cim. B}, 101:127, 1988.

\bibitem{Damour:1991rd}
T.~Damour and J.~H. Taylor.
\newblock {Strong Field Tests of Relativistic Gravity and Binary Pulsars}.
\newblock {\em Phys. Rev. D}, 45:1840--1868, 1992.

\bibitem{deRham:2012fw}
C.~de~Rham, A.~J. Tolley, and D.~H. Wesley.
\newblock {Vainshtein Mechanism in Binary Pulsars}.
\newblock {\em Phys. Rev. D}, 87(4):044025, 2013.

\bibitem{Desvignes:2019uxs}
G.~Desvignes, M.~Kramer, K.~Lee, J.~van Leeuwen, I.~Stairs, A.~Jessner,
  I.~Cognard, L.~Kasian, A.~Lyne, and B.~W. Stappers.
\newblock {Radio Emission from a Pulsar's Magnetic Pole Revealed by General
  Relativity}.
\newblock {\em Science}, 365(6457):1013--1017, 2019.

\bibitem{Dietrich:2020Sci}
T.~Dietrich, M.~W. Coughlin, P.~T.~H. Pang, M.~Bulla, J.~Heinzel, L.~Issa,
  I.~Tews, and S.~Antier.
\newblock {Multimessenger Constraints on the Neutron-Star Equation of State and
  the Hubble Constant}.
\newblock {\em Science}, 370(6523):1450--1453, 2020.

\bibitem{Doneva:2022ewd}
D.~D. Doneva, F.~M. Ramazano\u{g}lu, H.~O. Silva, T.~P. Sotiriou, and S.~S.
  Yazadjiev.
\newblock {Scalarization}.
\newblock {\em arXiv e-prints}, page arXiv:2211.01766, 2022.

\bibitem{Doneva:2017bvd}
D.~D. Doneva and S.~S. Yazadjiev.
\newblock {New Gauss-Bonnet Black Holes with Curvature-Induced Scalarization in
  Extended Scalar-Tensor Theories}.
\newblock {\em Phys. Rev. Lett.}, 120(13):131103, 2018.

\bibitem{Dong:2022zvh}
Y.~Dong, L.~Shao, Z.~Hu, X.~Miao, and Z.~Wang.
\newblock {Prospects for Constraining the Yukawa Gravity with Pulsars Around
  Sagittarius~A*}.
\newblock {\em JCAP}, 11:051, 2022.

\bibitem{1968Natur_Drake}
F.~D. Drake and H.~D. Craft.
\newblock {Second Periodic Pulsation in Pulsars}.
\newblock {\em Nature}, 220(5164):231--235, 1968.

\bibitem{Ferdman:2013xia}
R.~D. Ferdman et~al.
\newblock {The Double Pulsar: Evidence for Neutron Star Formation without an
  Iron Core-Collapse Supernova}.
\newblock {\em Astrophys. J.}, 767:85, 2013.

\bibitem{Fonseca:2021wxt}
E.~Fonseca et~al.
\newblock {Refined Mass and Geometric Measurements of the High-Mass PSR
  J0740+6620}.
\newblock {\em Astrophys. J. Lett.}, 915(1):L12, 2021.

\bibitem{Fradkin:1985ys}
E.~S. Fradkin and A.~A. Tseytlin.
\newblock {Quantum String Theory Effective Action}.
\newblock {\em Nucl. Phys. B}, 269:745, 1986.

\bibitem{Freire:2010mnras}
P.~C.~C. Freire and N.~Wex.
\newblock {The Orthometric Parameterisation of the Shapiro Delay and an
  Improved Test of General Relativity with Binary Pulsars}.
\newblock {\em MNRAS}, 409:199, 2010.

\bibitem{Freire:2012mg}
P.~C.~C. Freire, N.~Wex, G.~Esposito-Far{\`e}se, J.~P.~W. Verbiest, M.~Bailes,
  B.~A. Jacoby, M.~Kramer, I.~H. Stairs, J.~Antoniadis, and G.~H. Janssen.
\newblock {The Relativistic Pulsar-White Dwarf Binary PSR J1738+0333 II. The
  Most Stringent Test of Scalar-Tensor Gravity}.
\newblock {\em MNRAS}, 423:3328, 2012.

\bibitem{Gao:2021uus}
Y.~Gao, X.~Lai, L.~Shao, and R.~Xu.
\newblock {Rotation and Deformation of Strangeon Stars in the Lennard-Jones
  Model}.
\newblock {\em MNRAS}, 509(2):2758--2779, 2021.

\bibitem{Gao:2023mwu}
Y.~Gao, L.~Shao, and J.~Steinhoff.
\newblock {A Tight Universal Relation between the Shape Eccentricity and the
  Moment of Inertia for Rotating Neutron Stars}.
\newblock {\em Astrophys. J.}, 954(1):16, 2023.

\bibitem{Gold:1968Natur}
T.~Gold.
\newblock {Rotating Neutron Stars as the Origin of the Pulsating Radio
  Sources}.
\newblock {\em Nature}, 218:731--732, 1968.

\bibitem{Guo:2021leu}
M.~Guo, J.~Zhao, and L.~Shao.
\newblock {Extended reduced-order surrogate models for scalar-tensor gravity in
  the strong field and applications to binary pulsars and gravitational waves}.
\newblock {\em Phys. Rev. D}, 104(10):104065, 2021.

\bibitem{Hadzic:2020smr}
M.~Hadzic and Z.~Lin.
\newblock {Turning Point Principle for Relativistic Stars}.
\newblock {\em Commun. Math. Phys.}, 387(2):729--759, 2021.

\bibitem{2003Nature_hankins}
T.~H. Hankins, J.~S. Kern, J.~C. Weatherall, and J.~A. Eilek.
\newblock {Nanosecond Radio Bursts from Strong Plasma Turbulence in the Crab
  Pulsar}.
\newblock {\em Nature}, 422(6928):141--143, 2003.

\bibitem{Hartle:1967he}
J.~B. Hartle.
\newblock {Slowly Rotating Relativistic Stars. I. Equations of Structure}.
\newblock {\em Astrophys. J.}, 150:1005--1029, 1967.

\bibitem{Hewish:1968Nature}
A.~Hewish, S.~J. Bell, J.~D.~H. Pilkington, P.~F. Scott, and R.~A. Collins.
\newblock {Observation of a Rapidly Pulsating Radio Source}.
\newblock {\em Nature}, 217:709--713, 1968.

\bibitem{Hinderer:2007mb}
T.~Hinderer.
\newblock {Tidal Love Numbers of Neutron Stars}.
\newblock {\em Astrophys. J.}, 677:1216--1220, 2008.

\bibitem{Hinterbichler:2011tt}
K.~Hinterbichler.
\newblock {Theoretical Aspects of Massive Gravity}.
\newblock {\em Rev. Mod. Phys.}, 84:671--710, 2012.

\bibitem{Hobbs:2015}
G.~Hobbs.
\newblock {Developing a Pulsar-Based Time Standard}.
\newblock {\em Highlights Astron.}, 16:207--208, 2015.

\bibitem{Hu:2020ubl}
H.~Hu, M.~Kramer, N.~Wex, D.~J. Champion, and M.~S. Kehl.
\newblock {Constraining the Dense Matter Equation-of-State with Radio Pulsars}.
\newblock {\em MNRAS}, 497(3):3118--3130, 2020.

\bibitem{Hu:2021tyw}
Z.~Hu, Y.~Gao, R.~Xu, and L.~Shao.
\newblock {Scalarized Neutron Stars in Massive Scalar-Tensor Gravity: X-ray
  Pulsars and Tidal Deformability}.
\newblock {\em Phys. Rev. D}, 104(10):104014, 2021.

\bibitem{Hulse:1975ApJ}
R.~A. Hulse and J.~H. Taylor.
\newblock {Discovery of a Pulsar in a Binary System}.
\newblock {\em Astrophys. J. Lett.}, 195:L51--L53, 1975.

\bibitem{Jain_2010}
B.~Jain and J.~Khoury.
\newblock {Cosmological Tests of Gravity}.
\newblock {\em Annals Phys.}, 325:1479--1516, 2010.

\bibitem{Kramer:2022gru}
M.~Kramer.
\newblock {New Results from Testing Relativistic Gravity with Radio Pulsars}.
\newblock {\em Int. J. Mod. Phys. D}, 31(06):2230010, 2022.

\bibitem{Kramer:2006Sci}
M.~Kramer et~al.
\newblock {Tests of General Relativity from Timing the Double Pulsar}.
\newblock {\em Science}, 314:97--102, 2006.

\bibitem{Kramer:2021jcw}
M.~Kramer et~al.
\newblock {Strong-Field Gravity Tests with the Double Pulsar}.
\newblock {\em Phys. Rev. X}, 11(4):041050, 2021.

\bibitem{Kramer:2009zza}
M.~Kramer and N.~Wex.
\newblock {The Double Pulsar System: A Unique Laboratory for Gravity}.
\newblock {\em Class. Quant. Grav.}, 26:073001, 2009.

\bibitem{Lai:1993ve}
D.~Lai, F.~A. Rasio, and S.~L. Shapiro.
\newblock {Ellipsoidal Figures of Equilibrium-Compressible Models}.
\newblock {\em Astrophys. J. Suppl.}, 88:205--252, 1993.

\bibitem{Lai:2017ney}
X.~Lai and R.~Xu.
\newblock {Strangeon and Strangeon Star}.
\newblock {\em J. Phys. Conf. Ser.}, 861(1):012027, 2017.

\bibitem{Landry:2018ApJ}
P.~Landry and B.~Kumar.
\newblock {Constraints on the Moment of Inertia of PSR J0737-3039A from
  GW170817}.
\newblock {\em Astrophys. J. Lett.}, 868(2):L22, 2018.

\bibitem{Lattimer:2019Univ}
J.~M. Lattimer.
\newblock {Neutron Star Mass and Radius Measurements}.
\newblock {\em Universe}, 5(7):159, 2019.

\bibitem{Lattimer:2000nx}
J.~M. Lattimer and M.~Prakash.
\newblock {Neutron Star Structure and the Equation of State}.
\newblock {\em Astrophys. J.}, 550:426, 2001.

\bibitem{Li:2022qql}
H.~Li, Y.~Gao, L.~Shao, R.~Xu, and R.~Xu.
\newblock {Oscillation Modes and Gravitational Waves from Strangeon Stars}.
\newblock {\em MNRAS}, 516(4):6172--6179, 2022.

\bibitem{Lindblom:1992}
L.~Lindblom.
\newblock {Determining the Nuclear Equation of State from Neutron-Star Masses
  and Radii}.
\newblock {\em Astrophys. J.}, 398:569--573, 1992.

\bibitem{Liu:2014uka}
K.~Liu, R.~P. Eatough, N.~Wex, and M.~Kramer.
\newblock {Pulsar\textendash{}Black Hole Binaries: Prospects for New Gravity
  Tests with Future Radio Telescopes}.
\newblock {\em MNRAS}, 445(3):3115--3132, 2014.

\bibitem{Liu:2016bae}
K.~Liu et~al.
\newblock {Variability, polarimetry, and timing properties of single pulses
  from PSR J1713+0747 using the Large European Array for Pulsars}.
\newblock {\em Mon. Not. Roy. Astron. Soc.}, 463(3):3239--3248, 2016.

\bibitem{Liu:2011ae}
K.~Liu, N.~Wex, M.~Kramer, J.~M. Cordes, and T.~J.~W. Lazio.
\newblock {Prospects for Probing the Spacetime of Sgr A* with Pulsars}.
\newblock {\em Astrophys. J.}, 747:1, 2012.

\bibitem{Lorimer:2005LRR}
D.~R. Lorimer.
\newblock {Binary and Millisecond Pulsars}.
\newblock {\em Living Rev. Rel.}, 8:7, 2005.

\bibitem{2004handbook}
D.~R. Lorimer and M.~Kramer.
\newblock {\em {Handbook of Pulsar Astronomy}}.
\newblock Cambridge University Press, 2005.

\bibitem{Lu:2019gsr}
J.~Lu, K.~Lee, and R.~Xu.
\newblock {Advancing Pulsar Science with the FAST}.
\newblock {\em Sci. China Phys. Mech. Astron.}, 63(2):229531, 2020.

\bibitem{Lundgren:1995ApJ}
S.~C. Lundgren, J.~M. Cordes, M.~Ulmer, S.~M. Matz, S.~Lomatch, R.~S. Foster,
  and T.~Hankins.
\newblock {Giant Pulses from the Crab Pulsar: A Joint Radio and Gamma-Ray
  Study}.
\newblock {\em Astrophys. J.}, 453:433, 1995.

\bibitem{Lyne:2004Sci}
A.~G. Lyne et~al.
\newblock {A Double-Pulsar System: A Rare Laboratory for Relativistic Gravity
  and Plasma Physics}.
\newblock {\em Science}, 303:1153--1157, 2004.

\bibitem{Malhotra:1992}
R.~Malhotra, D.~Black, A.~Eck, and A.~Jackson.
\newblock {Resonant Orbital Evolution in the Putative Planetary System of
  PSR1257+12}.
\newblock {\em Nature}, 356(6370):583--585, 1992.

\bibitem{Manchester:2015mda}
R.~N. Manchester.
\newblock {Pulsars and Gravity}.
\newblock {\em Int. J. Mod. Phys. D}, 24(06):1530018, 2015.

\bibitem{Manchester:2005}
R.~N. Manchester, G.~B. Hobbs, A.~Teoh, and M.~Hobbs.
\newblock {The Australia Telescope National Facility Pulsar Catalogue}.
\newblock {\em Astron. J.}, 129:1993, 2005.

\bibitem{Mendes:2016fby}
R.~F.~P. Mendes and N.~Ortiz.
\newblock {Highly Compact Neutron Stars in Scalar-Tensor Theories of Gravity:
  Spontaneous Scalarization versus Gravitational Collapse}.
\newblock {\em Phys. Rev. D}, 93(12):124035, 2016.

\bibitem{Miao:2023gkr}
X.~L. Miao et~al.
\newblock {Variability, polarimetry, and timing properties of single pulses
  from PSR J2222\ensuremath{-}0137 using FAST}.
\newblock {\em Mon. Not. Roy. Astron. Soc.}, 526(2):2156--2166, 2023.

\bibitem{Miller:2019cac}
M.~C. Miller et~al.
\newblock {PSR J0030+0451 Mass and Radius from $NICER$ Data and Implications
  for the Properties of Neutron Star Matter}.
\newblock {\em Astrophys. J. Lett.}, 887(1):L24, 2019.

\bibitem{Napolitano:2012fp}
N.~R. Napolitano, S.~Capozziello, A.~J. Romanowsky, M.~Capaccioli, and
  C.~Tortora.
\newblock {Testing Yukawa-Like Potentials From f(R)-Gravity in Elliptical
  Galaxies}.
\newblock {\em Astrophys. J.}, 748:87, 2012.

\bibitem{Niebauer:1987ua}
T.~M. Niebauer, M.~P. Mchugh, and J.~E. Faller.
\newblock {Galilean Test for the Fifth Force}.
\newblock {\em Phys. Rev. Lett.}, 59:609--612, 1987.

\bibitem{Oppenheimer:1939ne}
J.~R. Oppenheimer and G.~M. Volkoff.
\newblock {On Massive Neutron Cores}.
\newblock {\em Phys. Rev.}, 55:374--381, 1939.

\bibitem{Peron:2014pba}
R.~Peron.
\newblock {Testing General Relativistic Predictions with the LAGEOS
  Satellites}.
\newblock {\em Adv. High Energy Phys.}, 2014:791367, 2014.

\bibitem{EventHorizonTelescope:2020qrl}
D.~Psaltis et~al.
\newblock {Gravitational Test Beyond the First Post-Newtonian Order with the
  Shadow of the M87 Black Hole}.
\newblock {\em Phys. Rev. Lett.}, 125(14):141104, 2020.

\bibitem{Ramazanoglu:2016kul}
F.~M. Ramazano\u{g}lu and F.~Pretorius.
\newblock {Spontaneous Scalarization with Massive Fields}.
\newblock {\em Phys. Rev. D}, 93(6):064005, 2016.

\bibitem{Ransom:2014}
S.~M. Ransom et~al.
\newblock {A millisecond pulsar in a stellar triple system}.
\newblock {\em Nature}, 505:520, 2014.

\bibitem{Reasenberg:1979ey}
R.~D. Reasenberg, I.~I. Shapiro, P.~E. MacNeil, R.~B. Goldstein, J.~C.
  Breidenthal, J.~P. Brenkle, D.~L. Cain, T.~M. Kaufman, T.~A. Komarek, and
  A.~I. Zygielbaum.
\newblock {Viking relativity experiment: Verification of signal retardation by
  solar gravity}.
\newblock {\em Astrophys. J. Lett.}, 234:L219--L221, 1979.

\bibitem{Reina:2015jia}
B.~Reina.
\newblock {Slowly Rotating Homogeneous Masses Revisited}.
\newblock {\em MNRAS}, 455(4):4512--4517, 2016.

\bibitem{Riley:2019yda}
T.~E. Riley et~al.
\newblock {A $NICER$ View of PSR J0030+0451: Millisecond Pulsar Parameter
  Estimation}.
\newblock {\em Astrophys. J. Lett.}, 887(1):L21, 2019.

\bibitem{Robertson1991}
D.~S. Robertson, W.~E. Carter, and W.~H. Dillinger.
\newblock {New Measurement of Solar Gravitational Deflection of Radio Signals
  Using VLBI}.
\newblock {\em Nature}, 349(6312):768--770, 1991.

\bibitem{Saffer:2021gak}
A.~Saffer and K.~Yagi.
\newblock {Tidal Deformabilities of Neutron Stars in Scalar-Gauss-Bonnet
  Gravity and their Applications to Multimessenger Tests of Gravity}.
\newblock {\em Phys. Rev. D}, 104(12):124052, 2021.

\bibitem{Sathyaprakash:2019yqt}
B.~S. Sathyaprakash et~al.
\newblock {Extreme gravity and fundamental physics}.
\newblock {\em Bull. Am. Astron. Soc.}, 51(3):251, May 2019.

\bibitem{Sennett:2017lcx}
N.~Sennett, L.~Shao, and J.~Steinhoff.
\newblock {Effective action model of dynamically scalarizing binary neutron
  stars}.
\newblock {\em Phys. Rev. D}, 96(8):084019, 2017.

\bibitem{Shannon:2012tr}
R.~M. Shannon and J.~M. Cordes.
\newblock {Pulse intensity modulation and the timing stability of millisecond
  pulsars: A case study of PSR J1713+0747}.
\newblock {\em Astrophys. J.}, 761:64, 2012.

\bibitem{Shao:2019gjj}
L.~Shao.
\newblock {Degeneracy in Studying the Supranuclear Equation of State and
  Modified Gravity with Neutron Stars}.
\newblock {\em AIP Conf. Proc.}, 2127(1):020016, 2019.

\bibitem{Shao:2022wdn}
L.~Shao.
\newblock {Imaging supermassive black hole shadows with a global very long
  baseline interferometry array}.
\newblock {\em Front. Phys. (Beijing)}, 17(4):44601, 2022.

\bibitem{Shao:2022izp}
L.~Shao.
\newblock {Radio Pulsars as a Laboratory for Strong-Field Gravity Tests}.
\newblock {\em Lect. Notes Phys.}, 1017:385--402, 2023.

\bibitem{Shao:2014wja}
L.~Shao et~al.
\newblock {Testing Gravity with Pulsars in the SKA Era}.
\newblock {\em PoS}, AASKA14:042, 2015.

\bibitem{Shao:2017gwu}
L.~Shao, N.~Sennett, A.~Buonanno, M.~Kramer, and N.~Wex.
\newblock {Constraining Nonperturbative Strong-Field Effects in Scalar-Tensor
  Gravity by Combining Pulsar Timing and Laser-Interferometer
  Gravitational-Wave Detectors}.
\newblock {\em Phys. Rev. X}, 7(4):041025, 2017.

\bibitem{Shao:2016ezh}
L.~Shao and N.~Wex.
\newblock {Tests of Gravitational Symmetries with Radio Pulsars}.
\newblock {\em Sci. China Phys. Mech. Astron.}, 59(9):699501, 2016.

\bibitem{Shao:2020fka}
L.~Shao, N.~Wex, and S.~Zhou.
\newblock {New Graviton Mass Bound from Binary Pulsars}.
\newblock {\em Phys. Rev. D}, 102(2):024069, 2020.

\bibitem{Shao:2022koz}
L.~Shao and K.~Yagi.
\newblock {Neutron Stars as Extreme Laboratories for Gravity Tests}.
\newblock {\em Sci. Bull.}, 67:1946--1949, 2022.

\bibitem{Silva:2020acr}
H.~O. Silva, A.~M. Holgado, A.~C\'ardenas-Avenda\~no, and N.~Yunes.
\newblock {Astrophysical and Theoretical Physics Implications from
  Multimessenger Neutron Star Observations}.
\newblock {\em Phys. Rev. Lett.}, 126(18):181101, 2021.

\bibitem{Silva:2017uqg}
H.~O. Silva, J.~Sakstein, L.~Gualtieri, T.~P. Sotiriou, and E.~Berti.
\newblock {Spontaneous Scalarization of Black Holes and Compact Stars from a
  Gauss-Bonnet Coupling}.
\newblock {\em Phys. Rev. Lett.}, 120(13):131104, 2018.

\bibitem{1968Sci_stealin}
D.~H. Staelin and I.~Reifenstein, Edward~C.
\newblock {Pulsating Radio Sources near the Crab Nebula}.
\newblock {\em Science}, 162(3861):1481--1483, 1968.

\bibitem{Stairs:2003eg}
I.~H. Stairs.
\newblock {Testing General Relativity with Pulsar Timing}.
\newblock {\em Living Rev. Rel.}, 6:5, 2003.

\bibitem{Talmadge:1988qz}
C.~Talmadge, J.~P. Berthias, R.~W. Hellings, and E.~M. Standish.
\newblock {Model Independent Constraints on Possible Modifications of Newtonian
  Gravity}.
\newblock {\em Phys. Rev. Lett.}, 61:1159--1162, 1988.

\bibitem{Tauris:2017ApJ}
T.~M. Tauris et~al.
\newblock {Formation of Double Neutron Star Systems}.
\newblock {\em Astrophys. J.}, 846(2):170, 2017.

\bibitem{1979Natur_Taylor}
J.~H. Taylor, L.~A. Fowler, and P.~M. McCulloch.
\newblock {Measurements of General Relativistic Effects in the Binary Pulsar
  PSR 1913+16}.
\newblock {\em Nature}, 277:437--440, 1979.

\bibitem{Tolman:1939jz}
R.~C. Tolman.
\newblock {Static Solutions of Einstein's Field Equations for Spheres of
  Fluid}.
\newblock {\em Phys. Rev.}, 55:364--373, 1939.

\bibitem{Voisin:2020lqi}
G.~Voisin, I.~Cognard, P.~C.~C. Freire, N.~Wex, L.~Guillemot, G.~Desvignes,
  M.~Kramer, and G.~Theureau.
\newblock {An improved test of the strong equivalence principle with the pulsar
  in a triple star system}.
\newblock {\em Astron. Astrophys.}, 638:A24, 2020.

\bibitem{2007MNRAS_Wang}
N.~Wang, R.~N. Manchester, and S.~Johnston.
\newblock {Pulsar Nulling and Mode Changing}.
\newblock {\em MNRAS}, 377:1383--1392, 2007.

\bibitem{Weinberg:2021exr}
S.~Weinberg.
\newblock {On the Development of Effective Field Theory}.
\newblock {\em Eur. Phys. J. H}, 46(1):6, 2021.

\bibitem{1913_2016ApJ}
J.~M. Weisberg and Y.~Huang.
\newblock {Relativistic Measurements from Timing the Binary Pulsar PSR
  B1913+16}.
\newblock {\em Astrophys. J.}, 829(1):55, 2016.

\bibitem{2006A_A_Weltevrede}
P.~Weltevrede, R.~T. Edwards, and B.~W. Stappers.
\newblock {The Subpulse Modulation Properties of Pulsars at 21 cm}.
\newblock {\em Astron. Astrophys.}, 445:243, 2006.

\bibitem{Weltman:2018zrl}
A.~Weltman et~al.
\newblock {Fundamental Physics with the Square Kilometre Array}.
\newblock {\em Publ. Astron. Soc. Austral.}, 37:e002, 2020.

\bibitem{Wex:2014nva}
N.~Wex.
\newblock {Testing Relativistic Gravity with Radio Pulsars}.
\newblock In S.~M. Kopeikin, editor, {\em {Frontiers in Relativistic Celestial
  Mechanics: Applications and Experiments}}, volume~2, page~39. Walter de
  Gruyter GmbH, Berlin/Boston, 2014.

\bibitem{Will:2018bme}
C.~M. Will.
\newblock {\em {Theory and Experiment in Gravitational Physics}}.
\newblock Cambridge University Press, 2018.

\bibitem{Williams:2005rv}
J.~G. Williams, S.~G. Turyshev, and D.~H. Boggs.
\newblock {Lunar laser Ranging Tests of the Equivalence Principle with the
  Earth and Moon}.
\newblock {\em Int. J. Mod. Phys. D}, 18:1129--1175, 2009.

\bibitem{Witten:1984rs}
E.~Witten.
\newblock {Cosmic Separation of Phases}.
\newblock {\em Phys. Rev. D}, 30:272--285, 1984.

\bibitem{Wolszczan:1992zg}
A.~Wolszczan and D.~A. Frail.
\newblock {A Planetary system around the millisecond pulsar PSR~1257+12}.
\newblock {\em Nature}, 355:145--147, 1992.

\bibitem{Xu:2003xe}
R.~Xu.
\newblock {Solid Quark Matter?}
\newblock {\em Astrophys. J. Lett.}, 596:L59--L62, 2003.

\bibitem{Xu:2020vbs}
R.~Xu, Y.~Gao, and L.~Shao.
\newblock {Strong-Field Effects in Massive Scalar-Tensor Gravity for Slowly
  Spinning Neutron Stars and Application to X-ray Pulsar Pulse Profiles}.
\newblock {\em Phys. Rev. D}, 102(6):064057, 2020.

\bibitem{Xu:2021kfh}
R.~Xu, Y.~Gao, and L.~Shao.
\newblock {Neutron Stars in Massive Scalar-Gauss-Bonnet Gravity: Spherical
  Structure and Time-Independent Perturbations}.
\newblock {\em Phys. Rev. D}, 105(2):024003, 2022.

\bibitem{Yagi:2013awa}
K.~Yagi and N.~Yunes.
\newblock {I-Love-Q Relations in Neutron Stars and their Applications to
  Astrophysics, Gravitational Waves and Fundamental Physics}.
\newblock {\em Phys. Rev. D}, 88(2):023009, 2013.

\bibitem{Yagi:2016bkt}
K.~Yagi and N.~Yunes.
\newblock {Approximate Universal Relations for Neutron Stars and Quark Stars}.
\newblock {\em Phys. Rept.}, 681:1--72, 2017.

\bibitem{Yao:2017ApJ}
J.~M. Yao, R.~N. Manchester, and N.~Wang.
\newblock {A New Electron-density Model for Estimation of Pulsar and FRB
  Distances}.
\newblock {\em Astrophys. J.}, 835(1):29.

\bibitem{Zakharov:2018cbj}
A.~F. Zakharov, P.~Jovanovi\'c, D.~Borka, and V.~Borka~Jovanovi\'c.
\newblock {Constraining the Range of Yukawa Gravity Interaction from S2 Star
  Orbits III: Improvement Expectations for Graviton Mass Bounds}.
\newblock {\em JCAP}, 04:050, 2018.

\bibitem{Zakharov:2016lzv}
A.~F. Zakharov, P.~Jovanovic, D.~Borka, and V.~B. Jovanovic.
\newblock {Constraining the Range of Yukawa Gravity Interaction from S2 Star
  Orbits II: Bounds on Graviton Mass}.
\newblock {\em JCAP}, 05:045, 2016.

\bibitem{Zhao:2022vig}
J.~Zhao, P.~C.~C. Freire, M.~Kramer, L.~Shao, and N.~Wex.
\newblock {Closing a Spontaneous-scalarization Window with Binary Pulsars}.
\newblock {\em Class. Quant. Grav.}, 39(11):11LT01, 2022.

\bibitem{Zhao:2019suc}
J.~Zhao, L.~Shao, Z.~Cao, and B.-Q. Ma.
\newblock {Reduced-order surrogate models for scalar-tensor gravity in the
  strong field regime and applications to binary pulsars and GW170817}.
\newblock {\em Phys. Rev. D}, 100(6):064034, 2019.

\bibitem{Zhou:2021tgo}
E.~Zhou, K.~Kiuchi, M.~Shibata, A.~Tsokaros, and K.~Uryu.
\newblock {Evolution of Equal Mass Binary Bare Quark Stars in Full General
  Relativity: Could a Supramassive Merger Remnant Experience Prompt Collapse?}
\newblock {\em Phys. Rev. D}, 106(10):103030, 2022.

\bibitem{Zhu:2018etc}
W.~W. Zhu et~al.
\newblock {Tests of Gravitational Symmetries with Pulsar Binary J1713+0747}.
\newblock {\em Mon. Not. Roy. Astron. Soc.}, 482(3):3249--3260, 2019.

\end{thebibliography}


\end{document}